\documentclass[twocolumn,prb,amsmath,amssymb,unsortedaddress,showpacs,floatfix,citeautoscript,footinbib]{revtex4-1}
\usepackage{latexsym,epsfig,graphicx,cleveref}
\usepackage{dcolumn}
\usepackage{comment,amsfonts,multirow,color,subfigure,bm,textcomp,ctable}
\usepackage{srcltx,soul}
\usepackage[usenames,dvipsnames,svgnames,table]{xcolor}
\usepackage{etoolbox}
\usepackage[colorlinks,urlcolor=blue,citecolor=blue]{hyperref}

\newcommand{\bra}[1]{\langle #1 |}
\newcommand{\ket}[1]{| #1 \rangle}

\newcommand{\bee}{\begin{equation}}
\newcommand{\ee}{\end{equation}}
\newcommand{\bma}{\begin{pmatrix}}
\newcommand{\ema}{\end{pmatrix}}

\newcommand{\bZ}{\mathbb{Z}}
\newcommand{\bI}{\mathbb{I}}

\newcommand{\ignore}[1]{}

\begin{document}

\title{Tuning between singlet, triplet, and mixed pairing states in an extended\\ Hubbard chain}
\author{Kuei Sun}
\affiliation {Department of Physics, University of Cincinnati,
Cincinnati, Ohio 45221-0011, USA}
\author{Ching-Kai Chiu}
\affiliation {Department of Physics and Astronomy, University of
British Columbia, Vancouver, British Columbia, Canada V6T 1Z1}
\affiliation {Department of Physics, University of Illinois at
Urbana-Champaign, Urbana, Illinois 61801-3080, USA}
\author{Hsiang-Hsuan Hung}
\affiliation {Department of Physics, University of Texas at
Austin, Austin, Texas 78712-1192, USA}
\author{Jiansheng Wu}
\affiliation {Department of Physics, Hong Kong University of
Science and Technology, Hong Kong} \affiliation {Department of
Physics, South University of Science and Technology of China,
Shenzhen, China}

\date{March 27, 2014}

\pacs{03.75.Ss, 71.10.Fd, 74.20.Rp, 74.78.Na}

\begin{abstract}
We study spin-half fermions in a one-dimensional extended Hubbard
chain at low filling. We identify three triplet and one singlet
pairing channels in the system, which are independently tunable as
a function of nearest-neighbor charge and spin interactions. In a
large-size system with translational invariance, we derive gap
equations for the corresponding pairing gaps and obtain a
Bogoliubov--de Gennes Hamiltonian with its non-trivial topology
determined by the interplay of these gaps. In an open-end system
with a fixed number of particles, we compute the exact many-body
ground state and identify the dominant pairing revealed by the
pair density matrix. Both cases show competition between the four
pairing states, resulting in broad regions for each of them and
relatively narrow regions for mixed-pairing states in the
parameter space. Our results enable the possibility of tuning a
nanowire between singlet and triplet pairing states without
breaking time-reversal or $SU(2)$ symmetry, accompanied by a
change in the system's topology.
\end{abstract}

\maketitle

\section{Introduction}
Cooper pairing\cite{Cooper56} is a key ingredient for exploring
condensation, superconductivity and superfluidity in interacting
many-fermion systems\cite{Leggett06}. In an electronic system,
phonon-mediated pairing between two electrons through a singlet
channel accounts for the onset of conventional superconductivity,
which is well described by the Bardeen-Cooper-Schrieffer (BCS)
theory invented over a half century ago\cite{Bardeen57}. Since
then, pairing mechanisms via different spin and orbital channels
have been extensively investigated, resulting early in successful
understanding of triplet pair superfluid phases in liquid
$^3$He\cite{Anderson61,Balian63,Anderson73,Leggett75,Wheatley75,Lee97,Leggett06}
or later in active studies on a variety of unconventional
superconductors such as
cuprates\cite{Sigrist91,Dagotto94,Tsuei00,Demler04,Lee06} and iron
pnictides\cite{Stewart11,Chubukov12,Seo08,Mereo09,Wang09,Wu,Hung12}
with singlet pairing order parameters as well as several
heavy-fermion compounds\cite{Joynt02,Pfleiderer09} and strontium
ruthenate Sr$_2$RuO$_4$\cite{Mackenzie03,Maeno12} with triplet
ones.

Multiple pairing effects enable the possibility of a transition
(or crossover) from one energetically favorable pairing state to
another as the system parameters change. In a triplet pairing
case, superfluid $^3$He can undergo a first-order phase transition
between an equal-spin-pairing state and a specific $^3P_0$
spin-orbit pairing state ($^3$He-$A$ and $B$ phases,
respectively)\cite{Leggett75,Leggett06}, as a function of
temperature and pressure. In a singlet pairing case, the BCS-type
superconductor or superfluid with a uniform pairing order
parameter can undergo a transition to a state with spatially
oscillatory ones in the presence of spin imbalance or magnetic
field, such as the Fulde-Ferrell-Larkin-Ovchinnikov
state\cite{FF,LO} with its experimental evidence in CeCoIn$_5$
\cite{Matsuda07,Pfleiderer09} and cold $^6$Li gases\cite{Liao10},
or the theoretically proposed $p$-orbital pair
condensate\cite{Zhang10}. In addition, several exotic transitions
between $d$-$(d+is)$-$s$\cite{Musaelian96,Khodas12},
$(p+ip)$-$p$\cite{Gurarie05} and $(p+ip)$-$f$\cite{Cheng10}
orbital pairing orders have also been theoretically discussed.
However, all these cases show the changes of the order parameters
only in the orbital or $z$-component spin space, while the total
spin of the pairing order remains the same (singlet or triplet)
upon the transitions. A transition or crossover between singlet
and triplet pairing states was less studied.

Moreover, in three dimensions there is an interesting state
showing the coexistence of $s$- and $p$-wave pairing orders
(reminiscent of a fragmented condensate), provided the
interparticle potentials in triplet and singlet channels are both
energetically favorable\cite{Balian63}. Such a mixed state
survives merely in a restrictive parameter regime and has not been
much focused\cite{Leggett75}. In two dimensions, the mixed state
has been proposed with the assistance of spin-orbit
couplings\cite{Gorkov01}, interfacial barriers,\cite{Romano13} or
deformation in the Fermi surface.\cite{Kuboki01} Recent findings
have suggested a feasible proposal for this mixture, which is
proximity-induced $p$-wave superconductivity in a
ferromagnets/$s$-wave superconductor heterostructure
\cite{Volkov03,Bergeret05,Keizer06,Eschrig08,Linder10,Almog11,Klose12,Quarterman12,Leksin12,Bergeret13,Hikino13}.
In these devices, even if the competition between singlet and
triplet pairing orders always exists since the attractive
interaction between opposite spins accompanies with the desired
attractive interaction between same spins, they can coexist within
a range across the interface, with thickness comparable to the
superconducting coherence length. Nevertheless, the
ferromagnet/superconductor interface is strongly inhomogeneous
such that the mixed region can hardly be described as a uniform
phase. For effectively characterizing the quantum phases with
singlet, triplet, and mixed pairing order parameters, a
well-defined uniform system and its modeling ought to be further
investigated.

Recently, a lot of interest has been stimulated in one-dimensional
(1D) superconductors for their topological nontrivial properties
and potential application on quantum information
processing\cite{Kitaev01,Wilczek09,Franz10,Alicea12,Beenakker13}.
In a system of spinless fermions on an open chain, the
superconducting state, which has a $p$-wave (triplet) pairing
order parameter, has been shown in a given parameter range as a
topological state that carries one Majorana fermion on each end of
the chain\cite{Kitaev01}. In a case of spin-half fermions, the
Majorana fermion states can emerge within a heterostructure in the
presence of $s$-wave (singlet) pairing order, spin-orbit coupling
and magnetic Zeeman field\cite{Lutchyn10,Oreg10,Stoudenmire11},
which has been experimentally realized in semiconductor nanowires
having a proximity-induced superconducting
gap\cite{Mourik12,Deng12,Rokhinson12,Das12,Finck13,Churchill13}.
On the other hand, a singlet superconductor without spin-orbit and
magnetic couplings is always topologically trivial. Therefore,
regarding the equivalence between a spin-half system and two
copies of spinless ones in the limit of spin decoupling, one could
expect that the tuning between one-dimensional singlet and triplet
pairing states may induce a change in the system's topology and
hence provide a new route for topological manipulation. In
addition, the topological property of a mixed pairing state would
also be an interesting subject. From this point of view, systems
with inside tunable pairing channels would be more appropriate for
investigation.

In this paper, we study an extended one-dimensional Hubbard model
with nearest-neighbor charge and spin interactions, particularly
focusing on the pairing phenomena in uniform and low-filling
regimes. We show that the system contains all four possible
pairing channels in the pair spin space, with coupling strengths
that can be independently varied by the tuning of charge and spin
interactions. We apply a mean-field treatment on a large-size case
with translational invariance and will derive gap equations
characterizing two intraspin triplet, one interspin triplet as
well as one singlet pairing orders, and mixed regions of them. We
shall obtain the effective Bogoliubov--de Gennes (BdG) Hamiltonian
of the model and discuss its topological properties. Beyond the
mean-field treatment, we perform exact diagonalization on an
open-end chain with a fixed number of particles, with
modifications to reduce finite-size effects (see detailed
discussions in Sec.~\ref{sec:ED_a}). We compute pair fractions of
the exact many-body ground state that indicates dominant and
stable pair species toward large-size and low-filling regimes
(reminiscent of a pair condensate). The results will show a change
of dominant pair species from one to another as the corresponding
couplings vary, accompanied with a characteristic behavior of pair
susceptibility or entanglement entropy. The mixed pairing state
will also be identified in regions where more than one pair
species dominate. Finally we compare the mean-field and
exact-diagonalization results.

The paper is organized as follows. In Sec.~\ref{sec:Model} we
introduce the model Hamiltonian and phenomenologically discuss the
pairing physics in the system. In Sec.~\ref{sec:MF} we perform the
mean-field treatment on a translation-invariant system to derive
the gap equations, followed by discussions of the pairing behavior
as well as the topological properties of the system. In
Sec.~\ref{sec:ED} we compute the exact ground state of a
fixed-number open-end chain. We present data that show evolution
of dominant pair species as the function of couplings and plot
state diagrams that characterize various stable pairing states
including mixed ones. Finally we summarize this study in
Sec.~\ref{sec:Conclusion}.

\section{Model}\label{sec:Model}
In this section we introduce the model Hamiltonian and
phenomenologically discuss the pairing tendency in the system. We
begin with an extended 1D Hubbard Hamiltonian with charge as well
as spin interactions and represent it in a suggestive form that
directly pinpoints four independently tunable pairing channels. We
then write down a fixed-number BCS-type ansatz to explain how
various pair species energetically compete with each other.
Finally we discuss how the system's symmetry enables a mixed
pairing state.

The extended Hubbard model has a general form of
\begin{eqnarray}
\tilde H = \sum\limits_i \Big[ {\sum\limits_{\sigma  =  \uparrow ,
\downarrow } { - {{t}_\sigma }\left( {\hat c_{\sigma i}^\dag
{{\hat c}_{\sigma i + 1}} + {\rm{H}}{\rm{.c}}{\rm{.}}} \right) -
{{\mu }_\sigma }{{\hat n}_{\sigma i}}}
}\nonumber\\
+ U{{\hat n}_{ \uparrow i}}{{\hat n}_{ \downarrow i}}  +
\sum\limits_{\sigma ,\sigma ' =  \uparrow , \downarrow } {{{
V}_{\sigma \sigma '}}{{\hat n}_{\sigma i}}{{\hat n}_{\sigma 'i +
1}}}  + 4 J{{{\bf{\hat S}}}_i} \cdot {{{\bf{\hat S}}}_{i + 1}}
\Big] \label{eqn:Ham1},
\end{eqnarray}
where $\hat c_{\sigma i}^\dag$ creates a fermion of spin $\sigma$
on site $i$, $\hat n_{\sigma i}= \hat c_{\sigma i}^\dag \hat
c_{\sigma i}$ is the number operator, ${\bf{\hat S}}_i=\hat
c_{\alpha i}^\dag {\vec{\sigma}}_{\alpha \beta} \hat c_{\beta
i}/2$ is the spin operator with ${\vec \sigma }=\{\sigma^x,
\sigma^y, \sigma^z \}$ being Pauli matrices, $t$ is the
nearest-neighbor tunneling strength, and $\mu$ is the chemical
potential. The couplings $U$, $V$ and $J$ represent the on-site
charge, nearest-neighbor charge, and spin interactions,
respectively. The parameters $t$, $\mu$, and $V$\cite{tuningV} are
taken as spin-dependent for the most general case (notice that
${V_{ \uparrow  \downarrow }} = {V_{ \downarrow  \uparrow }}$ is
required for most physical interactions).

In the following, we consider a case in which two spin species are
balanced and have the same single-particle spectrum, or $t_\sigma
\to t$ and $\mu_\sigma \to \mu$. We also focus on low-filling
regimes in which the double occupancies are dilute such that the
onsite repulsion can be treated as effective contributions to the
chemical potential in a Hartree approximation, ${{\hat n}_{
\uparrow i}}{{\hat n}_{ \downarrow i}} \to \left\langle {{{\hat
n}_{ \downarrow i}}} \right\rangle {{\hat n}_{ \uparrow i}} +
\left\langle {{{\hat n}_{ \uparrow i}}} \right\rangle {{\hat n}_{
\downarrow i}}$. However, the nearest-neighbor charge and spin
interactions account for intersite correlations that are essential
for the pairing behavior (as we will discuss later) and hence can
not be decoupled as single-site quantities. (We will show later in
this section that the physics of interest does not qualitatively
alter even incorporating the onsite interaction as its original
form in Eq.~(\ref{eqn:Ham1}), no matter whether it is repulsive or
attractive.) Therefore, with the approximation for the onsite
repulsion, one can pinpoint the pairing channels by rewriting the
Hamiltonian of Eq.~(\ref{eqn:Ham1}) in a suggestive form using two
intrapin triplet pair operators ${\hat b_{ \sigma ,i}^\dag = \hat
c_{ \sigma,i+1}^\dag \hat c_{ \sigma, i}^\dag }$ for
$\sigma=\uparrow,\downarrow$ as well as two interspin triplet and
singlet pair operators ${\hat b_{ \pm ,i}^\dag  = (\hat c_{
\downarrow ,i+1}^\dag \hat c_{ \uparrow , i}^\dag  \pm \hat c_{
\uparrow ,i+1}^\dag \hat c_{ \downarrow , i}^\dag })/\sqrt{2}$,
respectively, as
\begin{eqnarray}
\hat{H}=\sum\limits_i \big ( \hat{H}^{0}_i + \hat{H}^{\rm{I}}_i
\big ),  \label{eqn:Ham2}
\end{eqnarray}
with the non-interacting part,
\begin{eqnarray}
\hat{H}^{0}_i&=  {\sum\limits_{\sigma  =  \uparrow , \downarrow }
{  -t\left( {\hat c_{\sigma i}^\dag {{\hat c}_{\sigma i + 1}} +
{\rm{H}}{\rm{.c}}{\rm{.}}} \right) - \mu {{\hat n}_{\sigma i}}} },
\end{eqnarray}
and the interacting part,
\begin{eqnarray}
\hat{H}^{\rm{I}}_i&= \sum\limits_{\alpha  =
\uparrow,\downarrow,\pm} {{g_\alpha }\hat b^\dag_{\alpha, i} \hat
b_{\alpha, i} }.
\end{eqnarray}
Here the four pair couplings $g_{\uparrow,\downarrow,\pm}$ are
independently tunable via the tuning of the charge and spin
interactions ${V_{ \uparrow \uparrow }}$, ${V_{ \downarrow
\downarrow }}$, ${V_{ \uparrow \downarrow }}$ and $J$ in
Eq.~(\ref{eqn:Ham1}) as
\begin{eqnarray}
{g_{\uparrow ( \downarrow )}} &=&
{V_{\uparrow\uparrow(\downarrow\downarrow) }} + J ,\label{eqn:gupdown}\\
{g_ + } &=& {V_{ \uparrow  \downarrow }} + J, \label{eqn:g+}\\
{g_ - } &=& {V_{ \uparrow  \downarrow }} - 3J\label{eqn:g-}.
\end{eqnarray}

The Hamiltonian of Eq.~(\ref{eqn:Ham2}) conserves the total number
of each spin species $N_{\uparrow(\downarrow)}$. We
phenomenologically discuss the pairing tendency by applying a
generalized number-conserving BCS ansatz\cite{numberconserving} on
the many-body ground state in the momentum space $k$,
\begin{eqnarray}
{{\psi _{{\rm{BCS}}}} = \mathcal{A}{{\prod\limits_{\alpha  =
\uparrow,\downarrow , \pm } {\left( {\sum\limits_k {{f_{\alpha
,k}}\hat b_{\alpha ,k}^\dag } } \right)} }^{M_\alpha }}}{\left|
{{\rm{vac}}} \right\rangle }.\label{eqn:BCS}
\end{eqnarray}
Here the pair operators are defined in terms of
Fourier-transformed single-particle operators $\{\hat
c^\dag_{\alpha, k}\}$, as ${\hat b_{ \uparrow ( \downarrow
),k}^\dag = \hat c_{ \uparrow ( \downarrow ),k}^\dag \hat c_{
\uparrow ( \downarrow ), - k}^\dag }$ and ${\hat b_{ \pm ,k}^\dag
= \hat c_{ \downarrow ,k}^\dag \hat c_{ \uparrow , - k}^\dag  \pm
\hat c_{ \uparrow ,k}^\dag \hat c_{ \downarrow , - k}^\dag }$,
$f_{\alpha,k}$ is the amplitude for $b_{\alpha ,k}^\dag$, ${\left|
{{\rm{vac}}} \right\rangle }$ is the vacuum state, and
$\mathcal{A}$ is the normalization constant. The total numbers of
each pair species $M_\alpha$ are subject to number conservation
relations $2M_{ \uparrow ( \downarrow )} + M_+ + M_ - = {N_{
\uparrow ( \downarrow )}}$. Such constraints enable an energetic
competition between each pair species. From this point of view, we
expect the ground state with the favor (disfavor) of intraspin
triplet, interspin triplet or singlet pairing [or $M_\alpha$
dominates (diminishes)] if the corresponding coupling $g_\alpha$
is negative (positive) or attractive (repulsive). From
Eqs.~(\ref{eqn:gupdown})--(\ref{eqn:g-}) we note that the
attractive charge interaction (negative $V$) always benefits
pairing. The antiferromagnetic spin coupling (positive $J$) leads
to the favor of singlet pairing, as reminiscent of the singlet
($d$-wave) superconducting order in the two-dimensional $t$-$J$
model\cite{Lee06}, while the ferromagnetic coupling (negative $J$)
favors the triplet pairing, as reminiscent of the
proximity-induced $p$-wave superconducting order in
ferromagnet-superconductor
junctions\cite{Volkov03,Bergeret05,Keizer06,Eschrig08,Linder10,Almog11,Klose12,Gingrich12,Leksin12,Bergeret13,Hikino13}.
If one considers the onsite interaction $U$ as its original form
in Eq.~(\ref{eqn:Ham1}), it will energetically contribute only to
the singlet pair species. In this case, one could follow the same
discussion above for the energetic competition between different
pair species, except now the effect considered from the
nearest-neighbor singlet coupling $g_-$ should be replaced by a
combined effect of $g_-$ itself and $U$. Therefore, we do not
expect a qualitative change in the trend of pairing tendency by
incorporating the $U$ term, no matter whether it is attractive or
repulsive, and can thus stay with Eq.~(\ref{eqn:Ham2}) both for
simplicity and without the loss of generality. The ansatz of
Eq.~(\ref{eqn:BCS}) also tells that once a pair species is more
energetically favorable than the others, its total number tends to
maximize. Therefore, only one dominant pairing order is usually
expected in a number-conserving system, unless such trend is
protected by symmetries as discussed below.

The system possesses time-reversal symmetry if
$g_\uparrow=g_\downarrow$ and $SU(2)$ symmetry if
$g_\uparrow=g_\downarrow=g_+$. These symmetries insert a
sufficient condition of the coexistence of multiple triplet
pairing orders. For example, both intraspin pairing orders should
simultaneously emerge in the presence of time-reversal symmetry,
and together accompany the interspin triplet one in the presence
of $SU(2)$ symmetry. We note that the mixture of the two intraspin
pairing orders [e.g., $M_\uparrow=M_\downarrow\neq0$ and $M_\pm=0$
in Eq.~(\ref{eqn:BCS})] is different from the interspin triplet
pairing state (e.g., $M_+\neq 0$ and
$M_{\uparrow,\downarrow,-}=0$). The former is a fragmented state
(which has more than one dominant pair species), while the latter
is spin coherent and known as an equal spin pairing state [${\hat
b_{ + ,k}^\dag \to \hat c_{ \uparrow ,k}^\dag \hat c_{ \uparrow ,
- k}^\dag  + \hat c_{ \downarrow ,k}^\dag \hat c_{ \downarrow , -
k}^\dag }$ after an $SU(2)$ roration], analogous to the liquid
$^3$He-A phase~\cite{Leggett06}. In the limit of $g_\pm \to 0$,
the Hamiltonian of Eq.~(\ref{eqn:Ham2}) decouples to two
independent chains, each of which is described by Kitaev's
spinless fermionic model\cite{Kitaev01} in the presence of $U(1)$
symmetry breaking, capable of carrying Majorana fermions in a
topologically nontrivial state. Starting from this limit, our
model provides a route studying various couplings between such two
chains and their evolution toward the singlet pairing (topological
trivial) regime, hinting of a topological phase transition.
Finally, we remark that triplet and singlet orders can coexist
without breaking any of the symmetries discussed above. However,
even if they coexist, we expect the mixture in a relatively narrow
parameter range where the two pair species are energetically
compatible, outside which one order can always overcome the other
and become dominant. In Secs.~\ref{sec:MF} and \ref{sec:ED} we use
two different methods investigating the competition between the
four pairing orders as a function of the four couplings and
identifying the dominant regions for each pair species or their
mixture.

\section{Mean-field treatment on a large-size system}\label{sec:MF}
In this section, we establish a mean-field treatment for the
extended Hubbard Hamiltonian $\hat H$ in Eq.~(\ref{eqn:Ham2}) with
translational invariance (large-size limit) at zero temperature to
understand the possibility of triplet and singlet pairings. First,
we start from the exact quantum partition function and perform a
Hubbard-Stratonovich transformation with one singlet and three
triplet auxiliary bosonic fields. After the transformation, we
obtain an effective BdG Hamiltonian and turn to discuss  its
topology with the four pairings. Back to the main track, we derive
the gap equations of pairings and then find the parameter range
corresponding to the presence of pairing.

Before proceed, we comment that although the mean-field treatment
does not incorporate quantum fluctuations, which could be
essential for studying the 1D physics, it has been widely applied
to describe various 1D superconducting states both qualitatively
and quantitatively. For example, the mean-field
solutions\cite{Mizushima05,*Parish07,*Liu07,*Sun11,*Baksmaty11,*Sun12}
for 1D spin-imbalanced superconductors well match those obtained
from unbiased methods\cite{Orso07,*Feiguin07,*Bolech09} and agree
with experimental findings\cite{Liao10}. In Appendix
\ref{Richardson}, we consider another supportive example of 1D
superconducting systems, the Richardson
model\cite{Roman02,Dukelsky04}, and show the mean-field solution
consistent with the exact one for characterizing the
superconducting phase. Moreover, our BdG Hamiltonian, which
exhibits interesting topological properties as discussed below,
can be effectively applied on nano-wires with proximity-induced
superconducting
gaps\cite{Mourik12,Deng12,Rokhinson12,Das12,Finck13,Churchill13},
producing potential realization of tunable 1D topological
superconductors. Therefore, our mean-field study in this section
is not only valid to a certain extent but is also useful from both
theoretical and practical standpoints.

The quantum partition function of the system can be written as
\begin{eqnarray}
 Z=\int \prod_{i } \mathfrak{D}{\bf c}_{ i}\mathfrak{D}
{\bf c}_{ i}^\dagger e^{-\int_0^\beta d\tau [ {\bf c}^\dagger_{
i}\partial_\tau {\bf c}_{ i}+\hat{H}_i^0+\hat{H}_i^{\rm{I}}({\bf
c}_i,{\bf c}_i^\dagger) ] },
\end{eqnarray}
where ${\bf c}_i=(c_{\uparrow i},c_{\downarrow i})^T$. We
introduce four bosonic (scalar) auxiliary fields
$\rho_i=(\Delta_{\uparrow i},\Delta_{\downarrow i},\Delta_{+
i},\Delta_{- i})$ corresponding to pairing $b_{\uparrow i},\
b_{\downarrow i},\ b_{+i}$ and $b_{-i}$ respectively to perform a
Hubbard-Stratonovich transformation for the partition function
\begin{eqnarray}
Z=\int \big ( \prod_{p} \mathfrak{D}{\bf c}_p\mathfrak{D} {\bf
c}_p^\dagger \big ) \prod_i  d \rho_i d \rho_i^*e^{-\int_0^\beta
d\tau  ({\bf c}^\dagger_{ i}\partial_\tau {\bf c}_{ i}+ S_i)}
,\nonumber\\
\end{eqnarray}
where ${\bf c}_p=(c_{\uparrow p},c_{\downarrow p})^T$ and
\begin{eqnarray}
S_i=\sum_{\alpha=\uparrow,\downarrow,\pm}\left
[-\frac{\Delta_{\alpha i}^*\Delta_{\alpha i}}{g_\alpha}
+\Delta_{\alpha i} b^\dag_{\alpha i}+ {\rm{H.c.}} \right ]+
\hat{H}^{0}_i.
\end{eqnarray}
Although the action gains extra degrees of freedom from the
auxiliary fields ($b_{\alpha i}$), the effective Hamiltonian with
$b_{\alpha i}$ becomes integrable for ${\bf c}^\dagger_{\alpha i}$
and ${\bf c}_{\alpha i}$. Later, ${\bf c}^\dagger_{\alpha i}$ and
${\bf c}_{\alpha i}$ will be integrated out, and the pairing gaps
$\Delta_\alpha$ will be determined by finding the local extremum
of the action. Furthermore, understanding the expression of the
action in momentum space is necessary to compute the gap equation
in the following steps. Before performing Fourier transformation,
we assume the auxiliary fields to be translation invariant so the
site index $i$ can be neglected. In the momentum space, the
partition function with the translation-invariant auxiliary fields
is rewritten as
\begin{widetext}
\begin{eqnarray}
\mathbf{Z}= \int \big( \prod_p \mathfrak{D}{\bf c}_{
p}\mathfrak{D} {\bf c}_{ p}^\dagger \big ) d \rho d \rho^*
e^{-\int_0^\beta d\tau \Big
[L\sum_{\alpha=\uparrow,\downarrow,\pm}\frac{\Delta_\alpha^*\Delta_\alpha}{g_\alpha}+\sum_p({\bf
c}^\dagger_{p}\partial_\tau {\bf c}_{ p}+ \bf C_p^\dag
H_p^{\rm{BdG}}\bf C_p)  \Big ]},
\end{eqnarray}
up to a constant multiplier. Here $L$ is the total number of the
system sites,
\begin{eqnarray}
H^{\rm{BdG}}_p= \bma
\frac{-2t \cos p -\mu}{2} & 0 & i \sin p \Delta_\uparrow & \frac{-i \sin p \Delta_+ - \cos p \Delta_-}{\sqrt{2}} \\
0 & \frac{-2t \cos p -\mu}{2} & \frac{-i \sin p \Delta_+ + \cos p\Delta_-}{\sqrt{2}} & i \sin p \Delta_\downarrow \\
-i \sin p \Delta_\uparrow^* & \frac{i \sin p \Delta_+^* + \cos p \Delta_-^*}{\sqrt{2}} & \frac{2t \cos p +\mu}{2} & 0 \\
\frac{i \sin p \Delta_+^* - \cos p\Delta_-^*}{\sqrt{2}} & -i \sin
p \Delta_\downarrow^* & 0 & \frac{2t \cos p +\mu}{2} \label{BdG H
p} \ema,
\end{eqnarray}
\end{widetext}
and $\bf C_p= \bma c_{\uparrow p} & c_{\downarrow p} &
c^\dagger_{\uparrow -p} & c^\dagger_{\downarrow -p} \ema^T $ is a
vector describing particle and hole variables. The effective
Hamiltonian $H^{\rm{BdG}}_p$ is identified as the well-known BdG
Hamiltonian\cite{DeGennes66} describing superconducting systems
 in the momentum space. If all the triplet gaps
vanish $\Delta_\uparrow=\Delta_\downarrow=\Delta_+=0$,
$H^{\rm{BdG}}_p$ return to the BCS pairing case. If $\Delta_+ $
and $\Delta_-$ vanish, the system of $H^{\rm{BdG}}_p$ can be
treated as two decoupled Kitaev's 1D chains\cite{Kitaev01}, which
are time-reversal partners.

Let us return to the parent Hamiltonian in Eq.~(\ref{eqn:Ham2}).
Before the Hubbard-Stratonovich transformation, the parent
Hamiltonian shows that the system preserves time-reversal symmetry
given $g_\uparrow=g_\downarrow$. Although $U(1)$ symmetry is
broken after the transformation, the time-reversal symmetry should
be preserved in $H^{\rm{BdG}}_p$. For spin-half particles, the
time-reversal symmetry is defined as $\Theta=is_y K$ in the spin
space, where $K$ is the complex conjugation operator, such that
$c^\dagger_\uparrow \rightarrow -c^\dagger_\downarrow$ and
$c^\dagger_\downarrow \rightarrow c^\dagger_\uparrow$. Therefore,
in the hole basis, the time-reversal symmetry is still of the same
form.  To preserve time-reversal symmetry in $H^{\rm{BdG}}_p$, the
constraints of the pairing gaps must be imposed:
\begin{eqnarray}
\Delta_\uparrow=\Delta_\downarrow^*,\ \Delta_+=-\Delta^*_+,\
\Delta_-=\Delta^*_-. \label{Trestrict pair}
\end{eqnarray}
In general, because of the $U(1)$ symmetry breaking, the phase of
each pairing gap can be arbitrarily chosen by a $U(1)$ gauge
transformation. However, under arbitrary $U(1)$ transformation,
the constraints above no longer hold, and the definition of the
time-reversal operator $\Theta$ also changes. Hence, to avoid the
ambiguities of the unfixed pairings and the expression of
$\Theta$, we require the $U(1)$ gauge fixed once the
time-reversal-invariant constraints are imposed.

In the following, we turn to investigate the topological phases of
the $H_p^{\rm{BdG}}$. The BdG Hamiltonian, which possesses
particle and hole bases, automatically preserves particle-hole
symmetry with the corresponding symmetry operator $\Xi=\sigma_x
K$, which exchanges particle and hole. On the other hand, for a
spin-$1/2$ system, the time-reversal operators obeys $\Theta^2=-1$
so this system belongs to the class DIII, which exhibits $\bZ_2$
topological property in one dimension. To determine the topology
of the 1D chain, we first consider a simple case where
$\Delta_+=\Delta_-=0$. The BdG Hamiltonian becomes block
diagonalized and each block can be treated as a Kitaev 1D chain.
Hence, the system corresponds to two decoupled Kitaev 1D chains.
We expect that two Majorana modes arise at each end of the entire
1D non-trivial system. Kitaev\cite{Kitaev01} shows that the
nontrivial region is given by $|\mu|<2 t$. Now we recover nonzero
$\Delta_+$ and $\Delta_-$ to discuss the topology. In the absence
of all triplet pairings, the topological phase of the singlet
pairing superconductor is expected to be trivial. This 1D chain is
either nontrivial or trivial so the boundary between the two
phases is to be determined. The boundary is topological phase
transition points where the energy gap is closed. To find the
transition points, we write down the energy spectrum of
$H^{\rm{BdG}}_p$,
\begin{eqnarray}
4 E_{\pm}^2=\left(2t\cos p +\mu\right)^2+\left(\sin p |\Delta_t|
\pm \cos p |\Delta_-|\right)^2,
\end{eqnarray}
where
\begin{eqnarray}
|\Delta_t|^2=&|\Delta_\uparrow|^2+|\Delta_\downarrow|^2+|\Delta_+|^2.
\end{eqnarray}
When $E_{\pm}=0$, the transition occurs. That is,
${2t|\Delta_t|}/{\sqrt{|\Delta_-|^2+|\Delta_t|^2}}=|\mu|$ is the
boundary of the non-trivial region. Because $t>|\mu|$ is the
nontrivial region in the Kitaev model, the region can be extended
to
\begin{eqnarray}
\frac{2t|\Delta_t|}{\sqrt{|\Delta_-|^2+|\Delta_t|^2}}>|\mu|,
\label{eqn:topo region}
\end{eqnarray}
for our model. Here we see that the system is always topologically
trivial in a purely singlet pairing state ($\Delta_- \neq 0,
\Delta_t=0$) and has the maximum topologically nontrivial region
(the same region as in Kitaev's model) in a purely triplet pairing
state ($\Delta_-=0, \Delta_t \neq 0$). In a mixed pairing state
($\Delta_-\neq0, \Delta_t \neq 0$), the enhancement of the singlet
pairing strength shrinks the topologically nontrivial region,
which indicates a topological order as a result from the
competition between singlet and triplet pairings. Our finding also
enables a different route for realizing a topological transition
via the tuning of the singlet pairing $|\Delta_-|$, given $t$,
$\mu$ and the triplet pairing $|\Delta_t|$ (the three components
in Kitaev's model) all fixed. The rigorous derivation of the
topologically nontrivial region by computing $\bZ_2$ invariant is
provided in Appendix \ref{Z2 compute} for interested readers.

Now our focus is back on the partition function $\mathbf{Z}$ to
determine the values of the pairings. We integrate out all of  the
fermion operators $c^\dagger_{\beta p}$ and $c_{\beta p}$ in the
partition function
\begin{eqnarray}
\mathbf{Z}=\int d \rho d \rho^* e^{\beta
\sum_{\alpha=\uparrow,\downarrow,\pm}\frac{\Delta_\alpha^*\Delta_\alpha}{g_\alpha}+\frac{1}{2}\sum_{p,n}\ln
\det (G_\Delta^{-1}) },
\end{eqnarray}
where
\begin{eqnarray}
G_\Delta^{-1}=H^{\rm{BdG}}_p-i\omega_n \bI_{4\times 4},
\end{eqnarray}
and $\omega_n=\pi (2n +1 )\beta$ is the Matsubara frequency. To
obtain the equilibrium state (extremum of the free energy) of the
system, we take a variation of the action with respect to the
pairing gaps, which generates four gap equations,
\begin{eqnarray}
\frac{\Delta^*_\uparrow}{g_\uparrow} &=&
-\frac{\Delta^*_\uparrow}{2\beta L }\sum_{p, n}  \frac{\sin^2 p(\omega_n^2+T^2+2D_-)}{(\frac{\omega_n^2+T^2}{2})^2+(\omega_n^2+T^2)D_++|D_-|^2},\nonumber\\ \label{up gap}\\
\frac{\Delta^*_\downarrow}{g_\downarrow} &=&
-\frac{\Delta^*_\downarrow}{2\beta L }\sum_{p, n} \frac{\sin^2 p(\omega_n^2+T^2+2D_-)}{(\frac{\omega_n^2+T^2}{2})^2+(\omega_n^2+T^2)D_++|D_-|^2}, \nonumber\\ \label{down gap}\\
\frac{\Delta^*_+}{g_+} &=&
-\frac{\Delta^*_+}{2\beta L }\sum_{p, n}  \frac{\sin^2 p(\omega_n^2+T^2+2D_-)}{(\frac{\omega_n^2+T^2}{2})^2+(\omega_n^2+T^2)D_++|D_-|^2},  \nonumber\\\label{plus gap equation}\\
\frac{\Delta^*_-}{g_-} &=& -\frac{\Delta^*_-}{2\beta L }\sum_{p,
n} \frac{\cos^2
p(\omega_n^2+T^2+2D_-)}{(\frac{\omega_n^2+T^2}{2})^2+(\omega_n^2+T^2)D_++|D_-|^2},\nonumber\\
\label{minus gap equation}
\end{eqnarray}
where
\begin{eqnarray}
T&=&t\cos p +\mu, \\
D_\pm&=&\pm\cos^2 p |\Delta_-|^2 +\sin^2 p |\Delta_t|^2.
\end{eqnarray}
Since the strategy to solve these gap equations depends on the
symmetry properties of the triplet couplings, we first focus on
the $SU(2)$-symmetry-preserving case
($g_+=g_\uparrow=g_{\downarrow} $) and then extend the results to
the $SU(2)$-symmetry-breaking case ($g_+\neq
g_\uparrow=g_\downarrow$).

When $SU(2)$ symmetry is preserved, Eqs.~(\ref{up
gap})--(\ref{plus gap equation}) divided by their own pairings are
identical. Only two gap equations are involved in determining the
values of the pairings, which is similar to the
$SU(2)$-symmetry-breaking case. In the following, we solve these
two gap equations in Eq.~(\ref{minus gap equation}) and in the
same form of Eqs.~(\ref{up gap})--(\ref{plus gap equation}) at
zero temperature. Therefore, as $\beta \rightarrow \infty$,
$\frac{\sum_{\omega_n}}{\beta}\rightarrow
\int^{\infty}_{-\infty}\frac{d\omega}{2\pi}$ due to Matsubara
frequency $\omega_n=\pi(2n+1)/\beta$.  After the integration of
$\omega$, the gap equations are given by
\begin{eqnarray}
\frac{1}{g_\gamma} &=& \frac{1}{L }\sum_{p\geq 0} \sin^2 p \bigg[
\frac{1}{A_+}+\frac{1}{A_-}\nonumber\\ && + \left | \frac{\cos p
\Delta_-}{\sin p \Delta_t} \right |
\left(\frac{1}{A_+}-\frac{1}{A_-} \right) \bigg ]
\label{plus gap simple}\\
\frac{1}{g_-} &=&\frac{1}{2L }\sum_{p} \cos^2 p \bigg [
\frac{1}{A_+}+\frac{1}{A_-}\nonumber\\ && + \left |\frac{\sin p
\Delta_t}{\cos p \Delta_-} \right|
\left(\frac{1}{A_+}-\frac{1}{A_-} \right) \bigg ] , \label{minus
gap simple}
\end{eqnarray}
where  $g_\uparrow=g_\downarrow=g_+ \equiv g_\gamma$ and
\begin{eqnarray}
A_\pm= \sqrt{2D_{+} +T^2\pm 2|\sin 2p \Delta_t \Delta_-|}.
\end{eqnarray}
We note that given a set of the coupling constants, the gap
equations simultaneously determine only the two $SU(2)$ invariants
$|\Delta _t|$ and $|\Delta_-|$. In other words, the value of each
triplet pairing can not be determined separately. The reason is
that the mean-field pairings $\Delta_\uparrow,$
$\Delta_\downarrow$ and $\Delta_+$ are actually not individually
invariant under $SU(2)$ transformation as shown in Appendix
\ref{Z2 compute}.

\begin{figure}
\includegraphics[width=7cm]{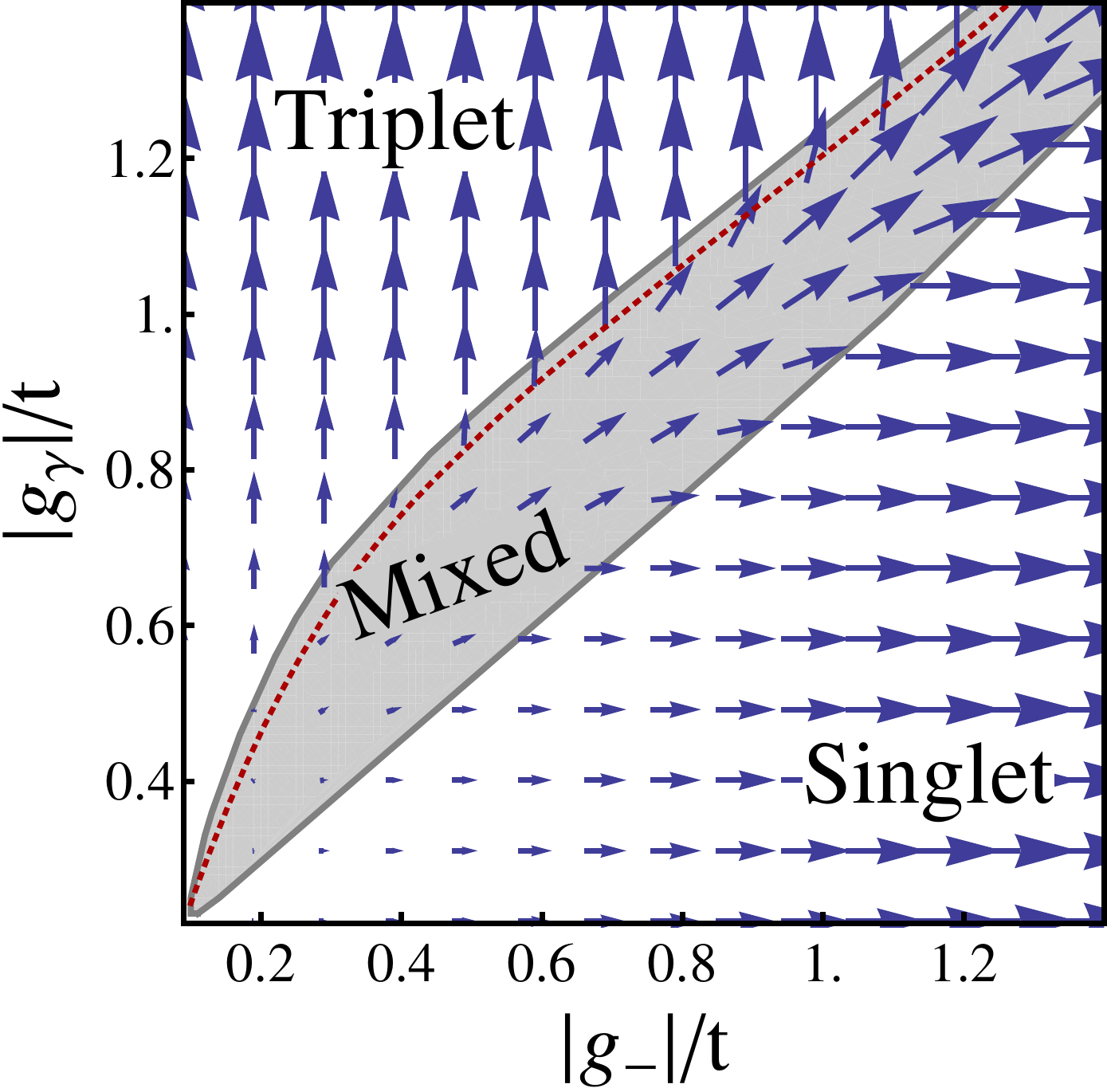}
    \caption{(Color online) Mean-field phase diagram characterizing
    singlet, triplet and mixed pairing states under time-reversal and $SU(2)$
    symmetries (when the three triplet couplings are equal, $g_\uparrow=g_\downarrow=g_+ \equiv g_\gamma$).
    The diagram is obtained by numerically solving the gap equations, which determine
    the equilibrium state of the system.
    Singlet ($|\Delta_-|$) and triplet ($|\Delta_t|$)
    pairing strengths are illustrated by vector
    arrows $(|\Delta_-|,|\Delta_t|)$ as a function of attractive singlet and
    triplet pair couplings, $g_-$ and $g_{\gamma}$, respectively
    (notice that both couplings are \emph{negative}). Each
    arrow has length proportional to
    $\sqrt{|\Delta_-|^2+|\Delta_t|^2}$ and slope equal to
    $|\Delta_-/\Delta_t|$. Purely triplet and purely singlet regions
    (filled with horizontal and vertical arrows, respectively)
    sandwich a relatively narrow mixed-pairing region
    (shadowed, filled with finite-slope arrows), with boundaries marked by gray solid lines.
    The red dashed line in the mixed region indicates
    the boundary between topologically trivial (below) and nontrivial (above) regions as
    the chemical potential $\mu=-1.7t$.
    There is no pairing beyond the left and bottom axes of this
    diagram.}
        \label{fig:f3_1} \vspace{-0.5cm}
\end{figure}

Numerically solving the gap equations in Eqs.~(\ref{plus gap
simple}) and (\ref{minus gap simple}) gives us the equilibrium
state of the system. We obtain a mixed pairing state (where
$|\Delta_-| \neq 0, |\Delta_t|\neq 0$) only in a restrictive
region in the parameter space of negative (attractive) $g_\gamma$
and $g_-$. Outside this region there is no mixed-pairing solution,
which means one or both of the gaps have to be zero. We thus solve
Eq.~(\ref{plus gap simple}) [Eq.~(\ref{minus gap simple})] for
$|\Delta_t|$ ($|\Delta_-|$) by setting $|\Delta_-|=0$
($|\Delta_t|=0$) in the triplet (singlet) coupling dominant region
$|g_\gamma| >|g_-|$ ($|g_-|>|g_\gamma|$). In Fig.~\ref{fig:f3_1}
we plot a phase diagram in the $|g_-|$-$|g_\gamma|$ plane for a
low-filling case of $\mu=-1.7t$ and draw a boundary (red dashed
curve) between topologically trivial and nontrivial regions. We
use vector arrows $(|\Delta_-|,|\Delta_t|)$ to represent singlet
and triplet pairing strengths, such that an arrow's length is
proportional to $\sqrt{|\Delta_-|^2+|\Delta_t|^2}$ and its slope
is equal to $|\Delta_-/\Delta_t|$. We see that the vector length
increases with the coupling strength. Horizontal and vertical
arrows indicate purely singlet and triplet pairing phases,
respectively, which sandwich a relatively narrow mixed-pairing
region of finite-slope arrows. There is no pairing in regions of
$|g_-|<0.1t$, $|g_\gamma|<0.2t$, or repulsive couplings. The
diagram agrees with the picture of energetic competition between
different pair species discussed in Sec.~\ref{sec:Model}; one can
imagine $g_-$ and $g_\gamma$ as two ``forces" that competitively
stretch and orient the vectors. Our data show that the arrow
smoothly rotates along a path from a singlet state to a triplet
one across the mixed region, implying a continuous evolution of
the system's free energy.

When $SU(2)$ symmetry is broken ($ g_+\neq
g_\uparrow=g_\downarrow$), the pairings $\Delta_+$ and
$\Delta_{\uparrow,\downarrow}$ are competing. Some pairings must
vanish to obey Eqs.~(\ref{up gap})--(\ref{plus gap equation}).
Determining the vanishing pairings involves the comparison of the
free energy corresponding to each pairing order. The one with
higher free energy should vanish. However, computing the free
energy is quite difficult. Instead, we give a qualitative argument
as we did in Sec.~\ref{sec:Model}. The negative values of the
coupling constants represent attractive interaction between the
electrons. From the energetic point of view, stronger attractive
force implies a higher possibility of pairing. Therefore, the
pairing with stronger attractive coupling wins the competition. We
can conclude that when $0\geq g_+>g_{\uparrow,\downarrow}$ ($0\geq
g_{\uparrow,\downarrow}> g_+$), $\Delta_+=0$
($\Delta_{\uparrow,\downarrow}=0$) and the pairings
$\Delta_{\uparrow,\downarrow}$ ($\Delta_+$) dominate. In this
case, Eqs.~(\ref{up gap})--(\ref{plus gap equation}) becomes
Eqs.~(\ref{plus gap simple}) and (\ref{minus gap simple}) with
$g_\gamma=g_{\uparrow,\downarrow}$ ($g_+$). As a result, the
survival pairings are also determined by Eqs.~(\ref{plus gap
simple}) and (\ref{minus gap simple}) and hence described by
Fig.~\ref{fig:f3_1}.

From the mean-field approach, the coupling constants control
singlet and triplet pairings. In the next section, we will study
the exact ground state of a fixed-number open-end chain and
compare the pairing behaviors with those in this section.

\section{Exact solutions of a fixed-number open-end system}\label{sec:ED}

In this section we perform exact diagonalization using the Lanczos
algorithm\cite{Lin90,Dagotto94} to solve the Hamiltonian of an
open-end chain with $L$ sites as well as fixed $N$ particles and
discuss the pairing physics showed by the results. The exact
solutions preserve all symmetries of the system and incorporate
effects of quantum fluctuations that are ignored in the mean-field
treatment. The $U(1)$ symmetry makes the Hamiltonian of
Eq.~(\ref{eqn:Ham2}) block-diagonalized with respect to the total
number of each spin species ($N_\uparrow$ and $N_\downarrow$, as
discussed in Sec.~\ref{sec:Model}) and hence allows us to deal
with only the block where the ground state locates. However, this
symmetry makes the original BCS-type pairing amplitude $\langle
\hat b_{\alpha}\rangle$ no longer a good order parameter for the
exact ground state.

Here we consider the pairing phenomenon as the condensation of
paired fermions\cite{Yang62,Leggett06}. To study this, one can
make an analogy to the condensation of bosons. In the Bose system,
a condensed state can be identified by macroscopic occupation of a
single-particle state, or mathematically, a macroscopic eigenvalue
of the single-particle density matrix\cite{Penrose56,Leggett06}.
In our Fermi system, it is the pair density matrix that is used to
identify the pairing as a trend toward the macroscopic occupation
of paired fermions. Specifically, we study the pairing tendency
(favor or disfavor of pairing) by comparing the largest eigenvalue
of the pair density matrix of the system with that of a free
system. The pair density matrix ${\rho ^{{\rm{pair}}}}$ is defined
as
\begin{eqnarray}
\rho _{{r_1}{\sigma _1},{r_2}{\sigma _2};{{r}_1'}{{\sigma
}_1'},{{r}_2'}{{\sigma }_2'}}^{{\rm{pair}}} = \left\langle {\hat
c_{{\sigma _1}{r_1}}^\dag \hat c_{{\sigma _2}{r_2}}^\dag {{\hat
c}_{{{\sigma}_2'}{{r}_2'}}}{{\hat c}_{{{\sigma}_1'}{{r}_1'}}}}
\right\rangle,\label{eqn:PDM}
\end{eqnarray}
where the matrix indices are denoted by a set of two-particle
states $\{ {r_1}{\sigma _1},{r_2}{\sigma _2}\}$ with $r$ and
$\sigma$ being spatial and spin quantum numbers, respectively. We
compute the eigen functions of ${\rho ^{{\rm{pair}}}}$ and find
that each of them is also an eigen state of a pair's total spin
${\hat {\bf S}^{{\rm{pair}}}}$ and its $z$ component $\hat
S_z^{{\rm{pair}}}$. Therefore, each eigen function falls into one
of the four pair classes including two intraspin triplet states
for $\uparrow/\downarrow$ ($\{
{S^{{\rm{pair}}}},S_z^{{\rm{pair}}}\} = \{ 1,\pm1\}$), one
interspin triplet state ($\{ 1,0\}$), and one singlet state ($\{
0,0\}$). From each class we find the largest eigen value
$\lambda^{(0)}$ and define a relative pair fraction as
\begin{eqnarray}
{P_\alpha } = \frac{{\lambda _\alpha ^{(0)} - 2}}{N}
\label{eqn:P},
\end{eqnarray}
where $\alpha=\uparrow,\downarrow,\pm$ denote the type of pairs in
the same convention as in Sec.~\ref{sec:Model} and $N=N_\uparrow
+N_\downarrow$ is the total number of particles. The relative pair
fraction $P_\alpha$ is evaluated as a comparison with a free
system, whose maximum eigenvalue is always
2~\cite{pairinfreesystem}. Since a free system has no pairing
preference, compared with this, positive (negative) $P_\alpha$
indicates the favor (disfavor) of $\alpha$ pair species. In the
thermodynamic limit, the onset of pair condensation is signaled by
$P \approx \lambda^{(0)} \sim O(1)$, although in most realistic
systems $P= 0.01\%$--$1\%$\cite{Leggett06}. In our case of an open
chain, we take $P$ (i) positive, (ii) increasing as the system
expands (by enlarging $L$ at fixed $N/L$), and (iii) increasing as
the system dilutes (by enlarging $L$ at fixed $N$) as three
signatures to identify a \emph{stable pairing state}. Signature
(ii) helps confirm the pairing tendency in the thermodynamic limit
(see the applications on the Richardson
model\cite{Roman02,Dukelsky04} and the original Hubbard model
discussed in Appendixes \ref{Richardson} and \ref{sec:HubbardU},
respectively), while (iii) does in the dilute regime of our
interest (see discussions in Sec.~\ref{sec:ED_b} and Appendix
\ref{sec:HubbardU}). Strictly speaking, such stable pairing state
of a finite-size chain is \emph{not} physically equivalent to a
pair condensate that should be defined in the thermodynamic limit
but could imply one if the trend persists. According to a theorem
in Ref.~[\onlinecite{Yang62}], the eigenvalues of a finite system
with $N$ fermions and $L$ sites are bounded as $\lambda^{(0)} \leq
N(2L-N+2)/2L$. Substituting a typical set in our calculations,
$L=20$ and $N=8$, we obtain $P \leq 60\%$.

In the following we focus on the time-reversal symmetric case, so
the number of independent couplings and hence that of independent
pair species is reduced by 1, allowing us to denote
$g_\uparrow=g_\downarrow \equiv g_\updownarrow$ and
$P_\uparrow=P_\downarrow \equiv P_\updownarrow$. In
Sec.~\ref{sec:ED_a} we discuss the finite-size effects and the
stability of pairing in the dilute limit. We suggest a
modification to maintain sufficient pairing tendency against the
finite-size effects without the lost of generality. In
Sec.~\ref{sec:ED_b}, we present results showing the evolution of
the system between different pairing states and the competition
between these pairings. We plot state diagrams characterizing
various stable pairing states as a function of couplings and
compare them with the mean-field results obtained in
Sec.~\ref{sec:MF}.

\subsection{Finite-size effects and stability of pairing}\label{sec:ED_a}

\begin{figure}[t]
\centering
\includegraphics[width=8.5cm]{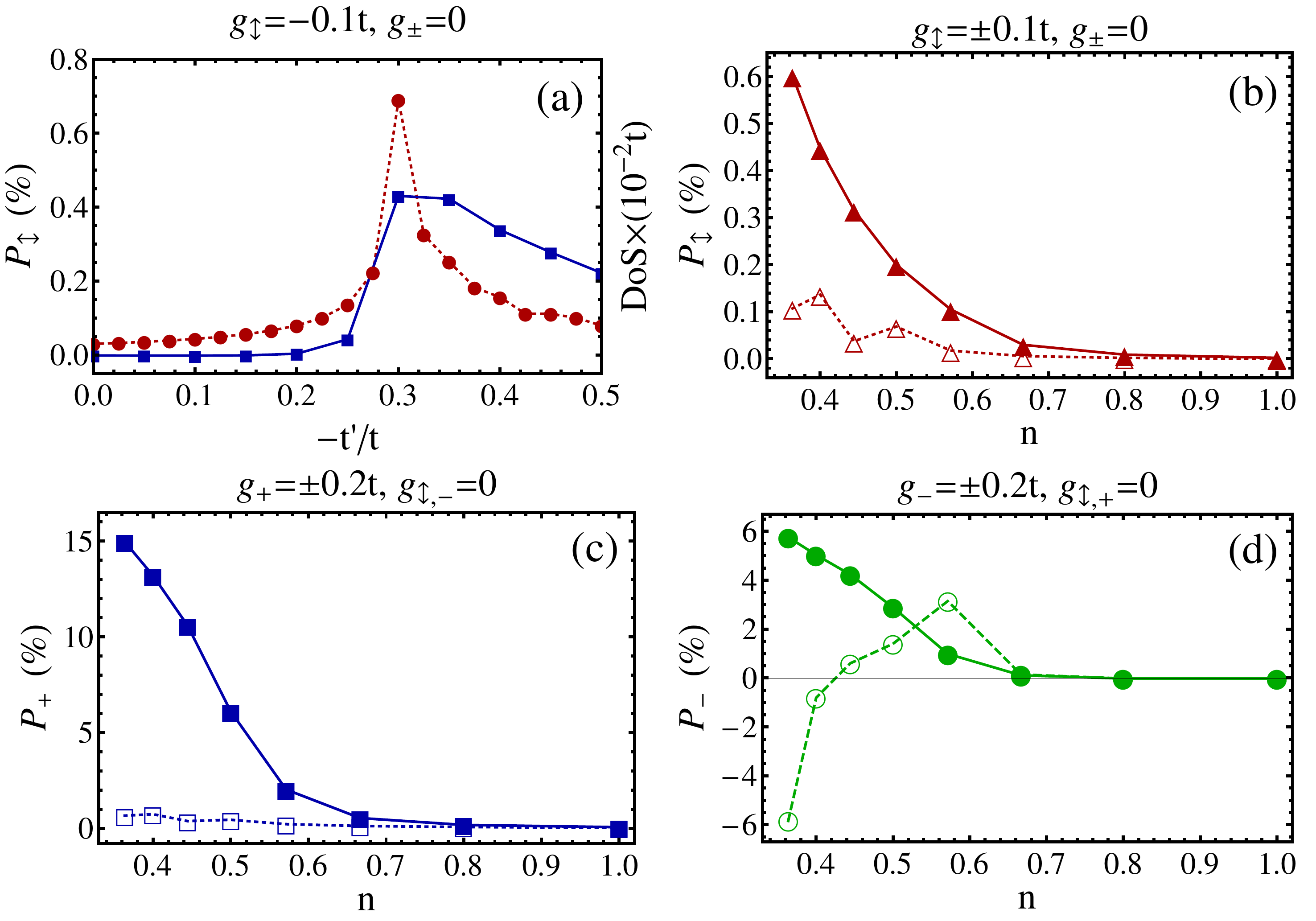}
       \caption{(Color online) (a) Intraspin pair fraction $P_\updownarrow$
       (solid curve, axis on the left of graph) and the single-particle density of states (DoS)
       at the Fermi level (dashed curve, axis on the right of graph)
       vs the second-nearest-neighbor tunneling $t'$.
       (b) Intraspin pair fraction $P_\updownarrow$ vs filling $n$
       at attractive ($g_\updownarrow=-0.1t$, solid curve) and repulsive ($0.1t$, dashed)
       pairing interactions while the other two couplings are set zero, $g_\pm=0$.
       (c),(d) Interspin triplet and singlet
       pair fractions $P_\pm$ vs $n$
       in attractive ($g_\pm=-0.2t$, respectively, solid curves) and repulsive ($0.2t$, dashed)
       cases, with the other two couplings set to zero as denoted in the plots.
       The relative pair fraction is measured from that of a free system, so negative values mean
       the disfavor of pairing.}
        \label{fig:f4_1} \vspace{-0.5cm}
\end{figure}

In a continuum system, only states within an energy scale of the
pairing gap around the Fermi level mainly participate in Cooper
pairing. In a finite-size chain of $L$ sites, the single particle
spectrum is always discrete and gapped by $O(t/L)$. At a weak
coupling of $|g_\alpha|<t/L$, it is the two degenerate states of
spin up and down at the Fermi level that mainly participate in the
interspin pairing, while the intraspin pairing is expected to be
more suppressed due to the lack of two such available states. In
fact, we explore the Hamiltonian of Eq.~(\ref{eqn:Ham2}) with
$N_{\uparrow,\downarrow}=4,L=8 \sim 24$ and find that $P_\pm>0$ in
a wide parameter range but $P_\updownarrow$ is always negative,
even in the range of $g_\updownarrow<0,|g_\updownarrow|\simeq t
\gg t/L$. In order to enhance the intraspin pairing, we increase
the single particle density of states around the Fermi level by
incorporating a second-nearest-neighbor tunneling into
Eq.~(\ref{eqn:Ham2}),
\begin{eqnarray}
\sum\limits_i {\sum\limits_{\sigma  =  \uparrow , \downarrow } { -
t'\left( {\hat c_{\sigma i}^\dag {{\hat c}_{\sigma i + 2}} +
{\rm{H}}{\rm{.c}}{\rm{.}}} \right)} }.  \label{eqn:Hamt2}
\end{eqnarray}
In Fig.~\ref{fig:f4_1}(a) we plot $P_\updownarrow$ (blue solid
curve) and the single-particle density of state at the Fermi level
(DoS, red dashed curve) as a function of the second
nearest-neighbor tunneling strength $t'$ for the case of an
attractive $g_\updownarrow=-0.1t$, $g_\pm=0$,
$N_\uparrow=N_\downarrow=4$, and $L=20$. We see that both
$P_\updownarrow$ and DoS increase as $t'$ increases from zero,
simultaneously reaching the maxima around $t'=-0.3t$. Such a trend
agrees with our expectation that the more states are around the
Fermi level, the higher pairing tendency the system shows. Below
we consider a combined Hamiltonian of Eqs.~(\ref{eqn:Ham2}) and
(\ref{eqn:Hamt2}) with $t'=-t/3$ so $P_\updownarrow$ is large and
positive ($\gg 0.01\%$) in a sufficiently large parameter regime.
Notice that we implement $t'$ to compensate the discreteness of
states due to the finite-size effects. In a large enough system,
we expect DoS around the Fermi level high enough for significant
pairing even with only the nearest-neighbor tunneling as in
Eq.~(\ref{eqn:Ham2}).

Now we turn to discuss the stability of pairing in the low-filling
regime of our interests. In the mean-field treatment in
Sec.~\ref{sec:MF}, the pairing order vanishes if the corresponding
coupling is positive (repulsive). In an open chain, we find that
the relative pair fraction can be (slightly) positive in the
repulsive regime. We attribute this to a finite-size effect and
expect that attraction instead of repulsion is the relevant
coupling for stable pairing as the system approaches the
low-filling limit via expansion in size. Figures
\ref{fig:f4_1}(b)--\ref{fig:f4_1}(d) show the three relative pair
fractions $P_{\updownarrow,+,-}$ as a function of filling number
$n=(N_\uparrow+N_\downarrow)/L$ at the corresponding coupling been
attractive (solid curves) or repulsive (dashed ones),
respectively. In each panel, we set the corresponding repulsive
(attractive) interaction as $g_{\updownarrow,+,-}>0$ ($<0$) and
keep the other two pairing effects irrelevant by setting the
couplings to zero. The filling is varied by the tuning of $L$ at
fixed $N_\uparrow=N_\downarrow=4$. We see that the pairing
tendencies are inapparent at half filling ($n=1$) in all cases.
Away from it, all the attractive cases show a monotonically
increasing $P$ toward lower fillings, while in the repulsive cases
$P$ either alternates in small positive values or becomes negative
in the low-filling regime. We confirm two of the
stable-pairing-state signatures discussed at the beginning of
Sec.~\ref{sec:ED} as (i) $P$ positive and (ii) monotonically
increasing toward lower filling. Therefore, only the attractive
interactions sustain a stable pairing state, in agreement with the
mean field results in Sec.~\ref{sec:MF}. In Sec.~\ref{sec:ED_b} we
use these two plus (iii) the increase of the pair fraction upon
the system's expansion at fixed filling to identify the stable
pairing states and study the tuning between them in a general case
in which more than one coupling is nonzero.

\subsection{Results and discussions}\label{sec:ED_b}

\begin{figure*}[t]
\centering
\includegraphics[width=13.2cm]{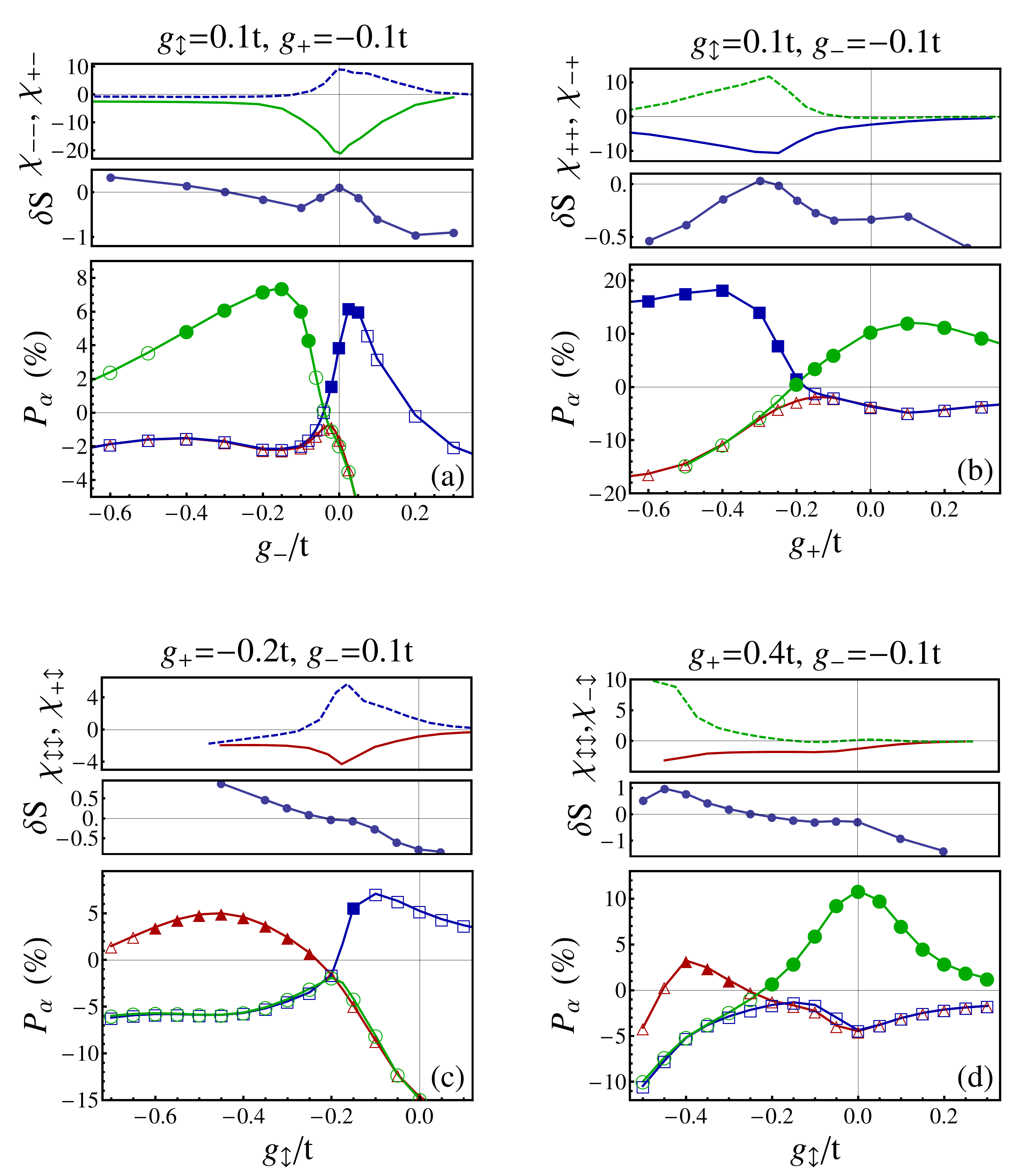}
       \caption{(Color online) Four cases shows relative pair
       fractions $P_\alpha$ (bottom panels), entanglement
       entropy $\delta S$ (middle) and pair susceptibility $\chi$ (top)
       tuned with the pairing couplings $g$. The intraspin, interspin
       triplet, and singlet pair fractions ($P_{\alpha=\updownarrow,+,-}$) are
       represented by red triangles, blue squares, and green circles, respectively. The
       filled (empty) symbols denote states that show (do not show) the three signatures
       for a \emph{stable pairing state} discussed in text.
       (a) Tuning between singlet and interspin triplet pairings as
       the singlet coupling $g_-$ varies, while the top panel shows
       pair susceptibilities $\chi_{--}$ (green solid curve) and $\chi_{+-}$ (blue dashed).
       (b) Tuning between interspin triplet and singlet pairings as
       the interspin triplet coupling $g_+$ varies, presented together with
       $\chi_{++}$ (blue solid) and $\chi_{-+}$ (green dashed).
       (c) Tuning between intraspin and interspin triplet pairings as
       the intraspin coupling $g_\updownarrow$ varies, presented together with
       $\chi_{\updownarrow \updownarrow}$ (red solid) and $\chi_{+ \updownarrow}$ (blue dashed).
       (d) Tuning between intraspin and singlet pairings as
       the intraspin coupling $g_\updownarrow$ varies, presented together with
       $\chi_{\updownarrow \updownarrow}$ (red solid) and $\chi_{-\updownarrow}$ (green dashed).
       }
        \label{fig:f4_2} \vspace{-0.5cm}
\end{figure*}

In this section, by computing the exact ground state of a
time-reversal symmetric open-end chain with $N=8$ and $L=20$ (thus
$N_\uparrow=N_\downarrow=4$ and the filling $N/L=0.4$) in a
sufficiently wide parameter range of $g_{\updownarrow,+,-}$, we
present results that show the evolution between different pairing
states and thus identify paths of tuning between singlet and
triplet or between multiple triplet pairing states in the
parameter space. We also obtain state diagrams characterizing the
stable regions for different pairing states. Following the three
signatures discussed at the beginning of Sec.~\ref{sec:ED}, a
stable pairing state of pair species $\alpha$ here is identified
by the relative pair fraction $P_\alpha$ (i) being positive, (ii)
increasing as compared with cases of $L=18$ and $L=16$ at fixed
$N=8$, and (iii) increasing as compared with that of $L=10$ at
fixed $N/L=0.4$. In addition, we calculate two other physical
quantities, pair susceptibility and von Neumann entanglement
entropy, and study their behaviors upon the cross between two
different stable-pairing regions. The pair susceptibility
$\chi_{\alpha \beta }$ is defined as a second derivative of the
ground-state energy $E_G$ with respect to the pairing couplings
$g_{\alpha}$ and $g_{\beta}$,
\begin{eqnarray}
{\chi _{\alpha \beta }} = \frac{{{\partial ^2}{E_G}}}{{\partial
{g_\alpha }\partial {g_\beta }}} \times t \label{eqn:chi},
\end{eqnarray}
with a multiplication of tunneling $t$ that makes $\chi$
dimensionless. According to the Hellmann-Feynman theorem, the
first derivative of $E_G$ with respect to $g_\alpha$, $\partial
{E_G}/\partial {g_\alpha } = \left\langle {\partial H/\partial
{g_\alpha }} \right\rangle$, is hence proportional to the total
number of $\alpha$ pairs on nearest-neighbor sites. Thus
$\chi_{\alpha \beta }$ describes the response of the total number
of such $\alpha$ pairs to $g_\beta$ (or $\beta$ pairs to
$g_\alpha$ since $\chi_{\alpha \beta }$=$\chi_{\beta \alpha }$).
The von Neumann entanglement entropy presented here is a relative
value measured from the free case (where all pairing couplings
vanish),
\begin{eqnarray}
\delta S =  - {\rm{Tr}}\left( {{\rho ^{{\rm{red}}}}\ln {\rho
^{{\rm{red}}}} - \rho _0^{{\rm{red}}}\ln \rho _0^{{\rm{red}}}}
\right), \label{eqn:entropy}
\end{eqnarray}
where ${{\rho ^{{\rm{red}}}}}$ is a reduced density matrix
constructed by tracing out the degrees of freedom of the
right-half chain, and ${{\rho_0 ^{{\rm{red}}}}}$ is that of a free
system. The relative entanglement entropy quantifies how much more
or less entangled (positive or negative $\delta S$, respectively)
the system is driven by the pairing couplings.

In Fig.~\ref{fig:f4_2}, we plot $P_{\updownarrow,+,-}$ (red
triangles, blue squares, and green circles, respectively) vs $g$
in four cases that show the tuning between different stable
pairing states (filled symbols in the $P$ curve contract to the
empty ones denoting states that do not satisfy the three stability
criterions). The bottom panel of Fig.~\ref{fig:f4_2}(a) shows the
tuning between interspin triplet and singlet pairing states ($P_+$
and $P_-$ dominates, respectively) as we vary $g_-$ and keep
$g_\updownarrow$ repulsive as well as $g_+$ attractive. We see
that the intraspin triplet pairing is always unfavorable
($P_\updownarrow<0$ everywhere). The interspin triplet pairing is
stable in a region of weakly positive and negative $g_-$, while
the interspin singlet pair fraction rises, overcomes the interspin
triplet one across a switch point where $P_+=P_-$, and becomes
stable as $g_-$ goes more negative. Toward the region of largely
positive (negative) $g_-$, the interspin triplet (singlet) pairing
decreases and becomes unstable. In the bottom panel of (b), we
plot the tuning between the same two pairing states but in a
different path in which $g_+$ is varied and $g_-$ is kept
attractive. We see a similar competition that the singlet pairing
dominates until is conquered by the interspin triplet one as $g_+$
goes sufficiently negative. The bottom panel of (c) [(d)] shows
how the stable intraspin triplet pairing state emerges with the
suppression of interspin triplet (singlet) pairing as
$g_\updownarrow$ varies from positive toward sufficiently negative
regions. In general, we find the tunability from stable
$\beta$-pairing to $\alpha$-pairing states, across a switch point
where $P_\alpha=P_\beta$, by varying $g_\alpha$ from positive to
sufficiently negative values and keeping $g_\beta$ a negative
constant, also in a condition that the other coupling $g_\gamma$
is set positive for the disfavor of $\gamma$ pairing all the time.
Both facts of (1) the switch between stable $\beta$- and
$\alpha$-pairing states around a negative $g_\alpha$ and (2)
increasing $P_\alpha$ accompanied with decreasing $P_\beta$ around
the switch point indicate a competition between the two pair
species: $g_\alpha$ has to overwhelm $g_\beta$ to make the
$\alpha$-pair species dominant. This results agrees with the
phenomenological discussions in Sec.~\ref{sec:Model} using the
number-conserving BCS ansatz of Eq.~(\ref{eqn:BCS}). The
competition also implies that a mixed state of two stable pairings
either hardly occurs or does so in a relatively small parameter
range. In fact, only in (b) do we see a mixture of weakly stable
interspin triplet and singlet pairings ($P\gtrsim 0$) around a
small region of $g_+=-0.2t$, while the other three cases lack such
mixture. We will discuss the mixed pairing state in more details
later.

Here we turn to study the pair susceptibility $\chi$, which could
show more information about the competition. The top panels of
(a)--(d), cases with tuning $g_\alpha$ at negatively constant
$g_\beta$, show $\chi_{\alpha \alpha}$ and $\chi_{\beta \alpha}$
(or the rate of change in numbers of nearest-neighbor $\alpha$ and
$\beta$ pairs with $g_\alpha$) vs $g_\alpha$ (solid and dashed
curves, respectively). We see in (a)--(c) that both $\chi_{\alpha
\alpha}$ and $\chi_{\beta \alpha}$ develop peaks with opposite
signs around the switch point where $P_\alpha=P_\beta$, reflecting
a drastic increase of $\beta$ pairs and drop of $\alpha$ pairs as
$g_\alpha$ increases toward the positive or repulsive region. The
slight mismatch between the switch point and the susceptibility
peaks can be due to the difference between $P$ and $\chi$; the
former represents pairs only for the dominant eigen wavefunction
of the pair density matrix, while the latter counts the
nearest-neighbor pairs only. In (d), neither
$\chi_{\updownarrow\updownarrow}$ nor $\chi_{-\updownarrow}$
exhibits a peak around the switch point $g_\updownarrow=-0.23t$.
This shows that the competition between intraspin triplet and
interspin singlet pairings is much weaker than that between any
other sets of two pairings. In addition, we plot the relative
entanglement entropy $\delta S$ vs $g$ on each of the middle panel
of (a)--(d). We see in most stable pairing regions in (a) and (b)
that the interspin triplet and singlet pairing states are less
entangled than the free system, or $\delta S<0$, while it reaches
a local maximum (slightly positive) around the switch point and
the peak of $\chi$. In the stable pairing regions in (c) and (d),
$\delta S$ monotonically decreases from positive to negative as
$g_\updownarrow$ increases, with its zero value exactly on the
switch point. These results show that the intraspin pairing state
tends to sustain higher long-range entanglement than the free
system, while the two interspin pairing ones do the opposite.

\begin{figure}[t]
\centering
   \includegraphics[width=8.5cm]{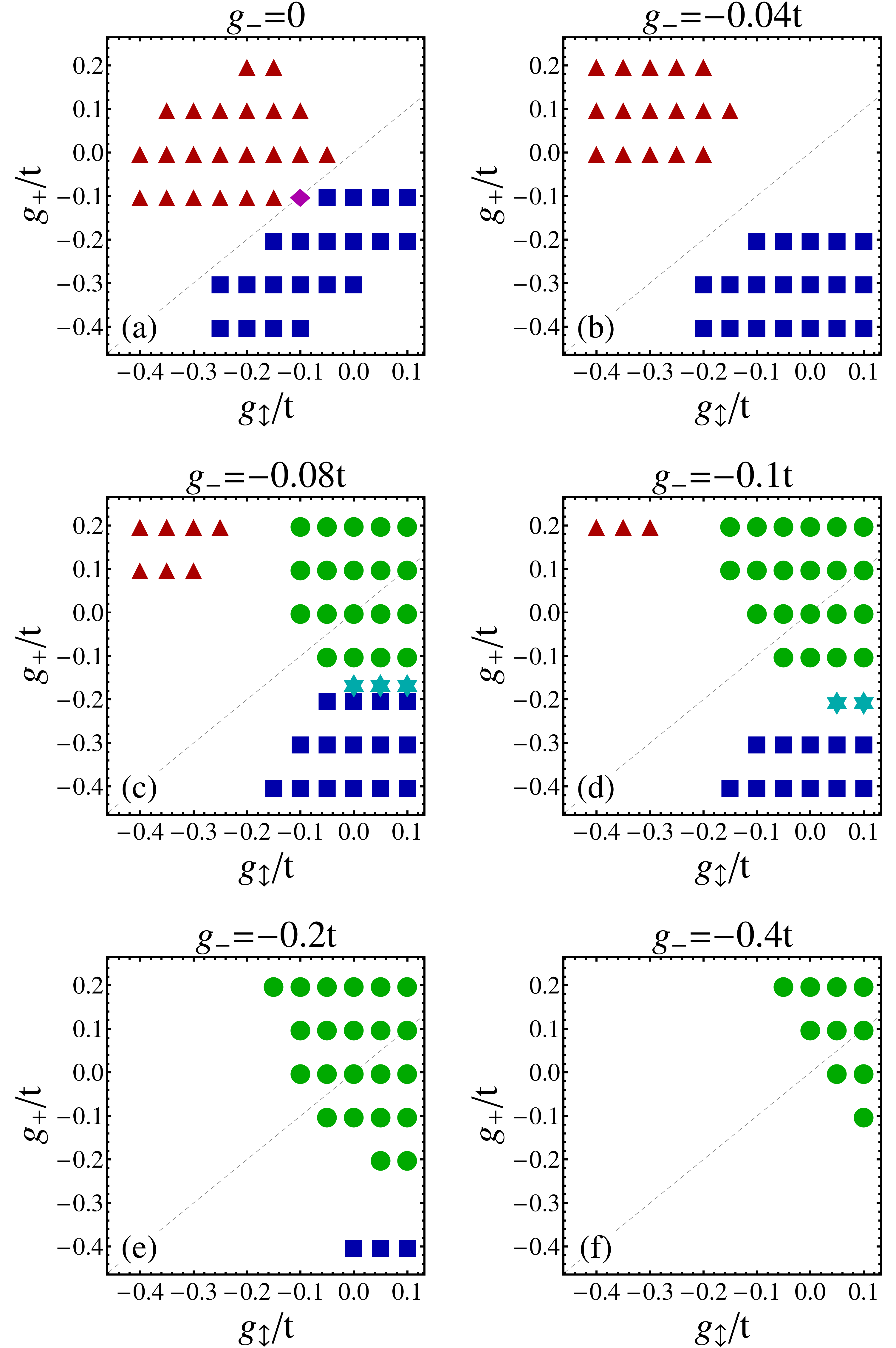}
       \caption{(Color online) (a)--(f)State diagrams showing stable pairing
       regions in $g_+$-$g_{\updownarrow}$ plane
       at $g_-/t=0$, $-0.04$, $-0.08$, $-0.1$, $-0.2$, and $-0.4$, respectively.
       Red triangles, blue squares and green circles represent intraspin, interspin
       triplet and singlet pairing, respectively, while the magenta diamonds and cyan stars
       represent a mix of intraspin and interspin pairings as well as that of interspin triplet and
       singlet pairings, respectively. The dashed lines $g_+=g_\updownarrow$ indicates $SU(2)$ symmetry
       of the system.}
        \label{fig:f4_3} \vspace{-0.5cm}
\end{figure}

In Fig.~\ref{fig:f4_3}, we plot state diagrams characterizing
regions of various stable paring states, including intraspin
triplet (denoted by triangles), interspin triplet (squares), and
singlet pairings (circles), as well as a mixture of the two
triplet pairings (diamonds) and that of the interspin triplet and
singlet pairings (stars), in the $g_\updownarrow$--$g_+$ plane at
a descending series of $g_-/t=0$, $-0.04$, $-0.08$, $-0.1$, $-0.2$
and $-0.4$ [(a)--(f), respectively]. The dashed line on each
diagram denotes the $SU(2)$-symmetric region where
$g_+=g_\updownarrow$. At $g_-=0$ [(a)], the diagram has stable
intraspin and interspin triplet pairing regions, which
qualitatively match $\{g_\updownarrow<0,g_+ > g_\updownarrow\}$
and $\{g_+<0,g_+ < g_\updownarrow\}$, respectively, indicating the
survival pairing state due to both the attractive interaction and
the success in competition against the other one. There is no
stable pairing state in a region where the two couplings are both
repulsive or both strongly attractive such that no one wins the
competition. The diagram also shows no stable singlet pairing
everywhere. Remarkably, we find a mixed pairing state with both
triplet pairings being stable on the overlap between the two
triplet pairing regions along the dashed line denoting $SU(2)$
symmetry. We check that the mixed state has the same pair
fractions of the two triplet pairings $P_\updownarrow=P_+$, in
agreement with the discussion in Sec.~\ref{sec:Model} that this
mixture is guaranteed by $SU(2)$ symmetry. [In fact, all data
points along the dashed lines in Fig.~\ref{fig:f4_3} show the same
set of eigenvalues corresponding to the intraspin and interspin
triplet pairings,
$\{\lambda^{(i)}_\updownarrow\}=\{\lambda^{(i)}_+\}$, reflecting
the $SU(2)$ symmetry of the pair density matrix (see details in
Appendix \ref{SU2}).] At $g_-=-0.04t$ [(b)], the two triplet
pairing regions separately move away from the dashed line, no
longer overlap, and hence leave no mixed pairing state. At
$g_-=-0.08t$ [(c)], the two triplet pairing regions further
separate and there appear singlet pairing states in the region of
positive or slightly negative $g_{\updownarrow,+}$. The singlet
pairing region overlaps the intraspin triplet one, producing a
mixed pairing region on a horizontal line of
$\{g_+=-0.16t=2g_-,g_\updownarrow \ge 0\}$. This mixture comprises
triplet and singlet pair species, which have different total spin
angular momentum but the same $\hat z$-component one. Since there
is no symmetry protection here, the pair fractions of both species
are not necessarily equal, or in general, $P_+ \neq P_-$. At
$g_-=-0.1t$ [(d)], the state diagram is similar to (c), with
further withdrawals of intraspin and interspin triplet pairing
regions toward the top-left and bottom-right corners,
respectively, an expansion of singlet pairing region, and a shift
of the mixed region of interspin triplet and singlet pairings to a
horizontal line of $\{g_+=-0.2t=2g_-,g_\updownarrow \ge 0.05t\}$.
At $g_-=-0.2t$ [(e)], the intraspin triplet pairing disappears in
the parameter range of interests, while the interspin triplet and
singlet pairing regions further separate from each other such that
the mixed region disappears as well. Finally, at a relatively
strong $g_-=-0.4t$ [(f)], only a small singlet pairing region
survives in the scope, occupying the top-right corner of the
diagram.

We turn to compare the mean-field results for a
translation-invariant system obtained in Sec.~\ref{sec:MF} and the
exact solutions for a fixed-number open-end chain here. First,
both cases show that a pairing state exists only if the
corresponding pairing coupling is attractive (negative). If two or
more pairing couplings are attractive, the corresponding pairing
states will compete with each other. Second, the quantities that
characterize pairing (the gaps in Sec.~\ref{sec:MF} or the pair
fractions here) always satisfy the same time-reversal or $SU(2)$
symmetry or both as the Hamiltonian does. Given time-reversal
symmetry, both cases can show mixed-pairing solutions of singlet
and triplet pairings. Given both time-reversal and $SU(2)$
symmetries, the mean-field case still shows this mixture but the
open-chain case does not. In addition, the open-chain case does
not exhibit notable topological signatures as the BdG Hamiltonian
does in the mean-field case. We attribute these issues to the
finite-size effects in the open-chain case and expect the two
cases' results closer to each other as the open-end chain size
increases. To achieve this, the study using density matrix
renormalization-group methods\cite{White93,Schollwock05} would be
helpful.

\section{Conclusion}\label{sec:Conclusion}
In this paper, we studied a low-filling Hubbard chain model with
nearest-neighbor charge and spin interactions, which produce four
independently tunable pairing couplings, corresponding to two
intrapin triplet, one interspin triplet, and one singlet pairing
channels, respectively. First, we performed a mean-field treatment
on a large-size system with translational invariance and derived
four gap equations characterizing the pairing order parameters.
The BdG Hamiltonian obtained in the treatment can exhibit
nontrivial topology in a chemical potential range that is the same
as Kitaev's model\cite{Kitaev01} in a purely triplet pairing state
but shrinks with the presence of a singlet pairing order. The
mean-field phase diagram under the time-reversal and $SU(2)$
symmetries shows a purely triplet or singlet pairing region if the
corresponding coupling overwhelms the other and a mixed pairing
region when both couplings are compatible. (After the completion
of this work, we perceived that two other works investigating
two-dimensional electronic systems also indicated a topological
phase transition due to the competition between triplet and
singlet pairing states.\cite{Lu13,Yao13}) Second, we employed an
exact-diagonalization algorithm to compute the many-body ground
state of an open-end fixed-number system with modification to
reduce the finite-size effect. We used three signatures of pair
fractions to identify a stable pairing state of the system, which
approaches a pair condensate if such trends persist. Our results
under the time-reversal symmetry show a stable intraspin triplet,
interspin triplet, or singlet pairing state in a region where the
corresponding coupling dominates and an overlapped region of mixed
intraspin and interspin triplet or mixed interspin triplet and
singlet pairing states. The system's switch from the singlet or
intraspin triplet pairing state to the interspin triplet one
accompanies a peak in the pair susceptibility, and that from the
singlet or interspin triplet pairing state to the intraspin
triplet one accompanies a sign change in the relative entanglement
entropy. Both the mean-field and exact-diagonalization cases
agreeably show a competitive nature of these pairings and hence
enable the tuning of the system between different pairing states
as well as mixtures of them.

Finally, we point out two platforms with properties suited for the
potential realization of tunable pairing channels---the key
mechanism in our model. First, recently focused Rydberg or
Rydberg-dressed atomic
gases\cite{Henkel10,*Pupillo10,*Saffman10,*Honer10,*Mukherjee11,*Schmidt-Kaler11,*Sevincli11,*Ji11,*Schaub12,*Viteau12,*Hague12,*Lauer12,*Robert-de-Saint-Vincent13,*Baluktsian13,*Mattioli13,*McQuillen13,*Honing13}
exhibit controllable $s$-wave and $p$-wave two-body
interactions\cite{Hamilton02,Kurz13} as well as significant
nearest-neighbor couplings when loaded in optical
lattices\cite{Pohl10,Weimer10,Viteau11,Anderson11,Saha14}. Second,
multispecies dipolar
gases\cite{Samokhin06,*Wu10,*Shi10,*LiaoR10,*Kain11,*Shi13,*Qi13}
have been investigated for the competition between short-range
singlet and long-range triplet interactions, capable of realizing
various pairing states and their mixture in higher-dimensional
systems. In addition, a recent experiment\cite{Greif13} has
demonstrated a method to measure the spin-correlation in optical
lattices, which is directly related to the pair fraction in our
study. However, how to tailor theses ideas to a practical scheme
for our chain lattices is a challenge. One of the future
directions is to study the model realization and to propose
experimental detection for its pairing order as well as
topological state.

\section*{Acknowledgments}
We are grateful to C. J. Bolech, Taylor L. Hughes, A. J. Leggett,
Shinsei Ryu, Nayana Shah and M. Stone for interesting discussions.
We acknowledge computational support from the Center for
Scientific Computing at the CNSI and MRL: NSF MRSEC (DMR-1121053)
and NSF CNS-0960316. This work was supported by DARPA-ARO Award
No. W911NF-07-1-0464 (KS), the University of Cincinnati (KS), the
Max Planck-UBC Center for Quantum Materials (CKC), the NSF
DMR-09-032991 (CKC), the HKRGC through Grant 605512, Grant 602813
and HKUST3/CRF09 (JW), and in part by Perimeter Institute for
Theoretical Physics (HHH). (Research at Perimeter Institute is
supported by the Government of Canada through Industry Canada and
by the Province of Ontario through the Ministry of Economic
Development and Innovation.) HHH and CKC thank the Department of
Physics at University of Cincinnati for the hospitality, where
part of the collaborative work took place.

\appendix

\section{Validity of mean-field and exact-diagonalization calculations on the Richardson model}\label{Richardson}

In this Appendix, we perform mean-field (MF) and
exact-diagonalization (ED) calculations on the Richardson
model\cite{Roman02,Dukelsky04}, which is a 1D exactly solvable
model approaching the BCS limit. Our results show the critical
coupling for the onset of superconductivity consistent with the
exact solution and thus the validity of both methods on 1D
superconducting systems (such as our model) to a certain extent.

For both $U(1)$-preserving finite-size and $U(1)$-breaking
infinite-size systems, computing the pair density matrix of
Eq.~(\ref{eqn:PDM}) is a valid method to determine the presence of
superconducting pairing\cite{Leggett06}. A macroscopic eigenvalue
of the pair density matrix shows the region of coupling constant
corresponding to a pair condensation or the superconducting
pairing. In the following, we calculate the pair density matrix by
performing ED on a few-body finite-size Richardson model and MF
treatment on the model in the thermodynamic limit. The Richardson
model is described by a half filling Hamiltonian in the form of
\begin{eqnarray}
H_{\rm{R}}=\frac{2}{1}\sum_{j=1,
\sigma=\uparrow,\downarrow}^N{\epsilon_{j\sigma}}
c^\dagger_{j\sigma}c_{j\sigma}-G\sum_{j,j'=1}^N
c^\dagger_{j\uparrow}c^\dagger_{j\downarrow}c_{j'\downarrow}c_{j'\uparrow},\nonumber\\
\end{eqnarray}
where $N$ denotes the number of sites, which is equal to the
number of particles, and $G$ is the coupling constant. The
Hamiltonian is different from the BCS\cite{Bardeen57} Hamiltonian
in lattices. The single-body term describes an on-site energy
($\epsilon_{j\sigma}$) instead of hopping, and the two-body term
represents interaction within all possible ranges rather than the
on-site one. The Hamiltonian above still preserves $U(1)$ symmetry
and can be exactly solved to obtain the many-body ground state and
the ground-state energy. By choosing some specific energy
$(\epsilon_{j\sigma})$ distribution, the physical phase of the
system can be determined in the thermodynamic limit ($N\rightarrow
\infty$). In the following, we discuss the onset of
superconductivity in a two-level distribution
$\epsilon_{j\sigma}=\pm\epsilon_1$. For the comparison between
different system sizes, we normalize the interacting effects by
defining a normalized coupling constant $g=GN$. We note that the
arrangement of the energies on each site does not change the
physical properties because the strength of the interaction in
each range is described by the same coupling constant $G$.

\begin{figure}
\includegraphics[width=7cm]{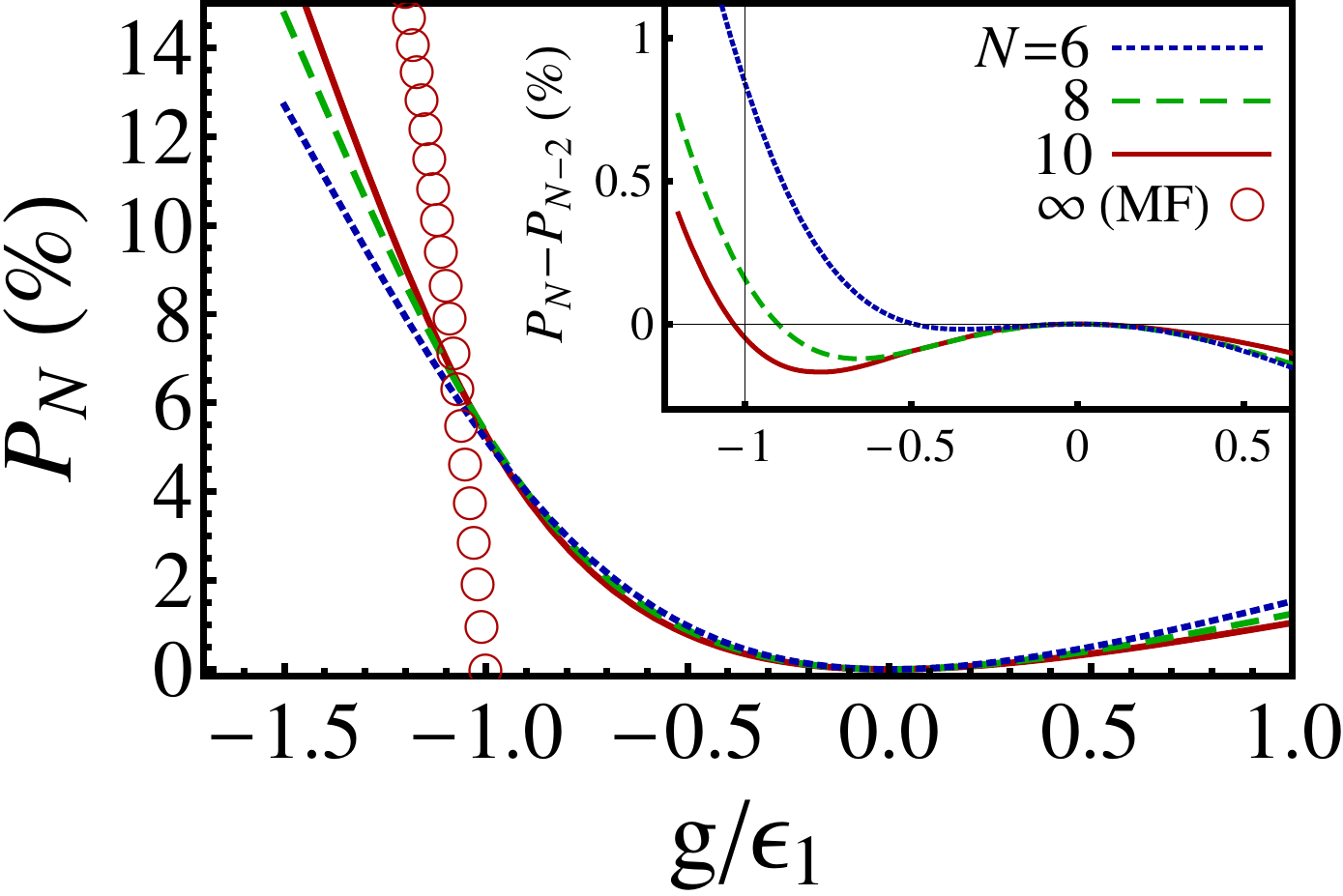}
\caption{(Color online) The relative pair fraction $P_N$ vs
normalized coupling constant $g$ at half-filling with three
different particle numbers of $N=6$, $8$, and $10$ (blue dotted,
green dashed, and red solid curves, respectively). In the region
of $g/\epsilon_1<-1$, larger $N$ means larger $P_N$, so the system
is regarded as a superconducting state. The inset shows that the
phase-transition point $P_{N}-P_{N-2}=0$ approaches the exact
solution $g_c=-\epsilon_1$ as $N$ increases. The red circles,
which are for an infinitely large-size system predicted by the
mean-field (MF) treatment, show that $P$ rises from zero exactly
at $g_c$.} \label{two level figure} \vspace{-0.5cm}
\end{figure}

By performing ED, we obtain the pair density matrices for the
ground states of $N=6$, $8$, and $10$. The presence of the
superconducting pairing is determined by the largest eigenvalue of
the pair density matrix being $O(N)$. In our ED case, the system
size is always too small to make a conclusion. Instead, we
calculate the relative pair fraction $P_N$ for $N$ particles as
defined in Eq.~(\ref{eqn:P}) but only for the singlet pairing here
(so the spin index $\alpha$ is dropped for convenience). The $-2$
in the definition of $P$ is to measure the eigenvalue from that of
a free system\cite{pairinfreesystem} (also see detailed
discussions in Sec.~\ref{sec:ED}). If the superconducting pairing
occurs, we expect that $P$ increases as $N$ increases, which
suggests that $P_N-P_{N-2}$ changes sign across the transition
point. As shown in the inset of Fig.~\ref{two level figure}, at
$N=10$ the transition point is near $g=-1$ and $P_N>P_{N-2}$ as
$g<-1$ so the region of $g<-1$ corresponds to possible
superconductor pairing. Our result is consistent with the
two-level Richardson model in the thermodynamic limit discussed in
Ref. \onlinecite{Roman02}.

In the following, we use a MF treatment to calculate the pair
fraction in the thermodynamic limit ($N \to \infty$) and compare
it with the results from ED as well as the exact solution. The
order parameter is defined as a function of spacial coordination
$\Delta_i=\sum_j
(-G_{ij})\left<c_{j\downarrow}c_{j\uparrow}\right>$ (where the
coupling $G_{ij}$ is first assumed spatial dependent). At the MF
level, the Hamiltonian can be rewritten as
\begin{eqnarray}
H^{\rm MF}_{\rm R}&=&\sum_i
\left(c^\dag_{i\uparrow},c_{i\downarrow}\right) H_i \left(
\begin{array}{c}
c_{i\uparrow}\\
c^\dag_{i\downarrow}
\end{array}
\right)-\sum_{ij}(G^{-1})_{ij}\Delta^*_{i}\Delta_j ,\nonumber\\
H_i &=& \left( \begin{array}{cc}
\frac{1}{2}(\epsilon_{i\uparrow}-\mu) & \Delta^*_i\\
\Delta_i & -\frac{1}{2}(\epsilon_{i\downarrow}-\mu)
\end{array}
\right).
\end{eqnarray}
The energy and the corresponding eigenstates are the followings,
\begin{eqnarray}
E_\pm&=&\mu_i\pm E_i,\\
\mu_i&=&\frac{1}{4}\left(\epsilon_{j\uparrow}-\epsilon_{j\downarrow} \right),\\
E_i &=&
\sqrt{\frac{1}{16}\left(\epsilon_{j\uparrow}+\epsilon_{j\downarrow}-2\mu\right)^2+|\Delta_j|^2} \nonumber \\
&\equiv& \sqrt{\varepsilon_i^2+|\Delta_i|^2},\\
\Phi^\dag_{i+} \left|{\psi _g}\right \rangle
&=& \left( \cos \frac{\theta_i}{2} c^\dag_{i\uparrow}+\sin\frac{\theta_i}{2}e^{i\phi_i} c_{i\downarrow}\right) \left|{\psi _g}\right \rangle \nonumber \\
&\equiv&  \left( u_i c^\dag_{i\uparrow}+v_i c_{i\downarrow}\right)
\left|{\psi _g}\right \rangle, \\
\Phi^\dag_{i-} \left|{\psi _g}\right \rangle
&=& \left( -\sin \frac{\theta_i}{2} c^\dag_{i\uparrow}+\cos\frac{\theta_i}{2}e^{i\phi_i} c_{i\downarrow}\right) \left|{\psi _g}\right \rangle,\\
\tan \theta_i &=& \frac{|\Delta_i|}{\varepsilon_i}, \tan
\phi_i=\frac{\rm Im \Delta_i}{\rm Re \Delta_i}.
\end{eqnarray}
Here $\left| {{\psi _g}} \right\rangle$ is the ground state and
$\Phi^\dag_{i\pm}$ are quasiparticle operators. Similar to the BCS
theory, we have the self-consistent gap equation as
\begin{eqnarray}
\Delta_i= \sum_j (-G_{ij}) \frac{\Delta_j}{2 E_j}\left[
\frac{1}{e^{\beta(\mu_j+E_j)}+1} -
\frac{1}{e^{\beta(\mu_j-E_j)}+1}\right].\nonumber\\
\end{eqnarray}
For simplicity, we consider the same setup as in the ED case,
$\mu_i=0$ and $\varepsilon_i=\epsilon_1/2$. By assuming the
homogeneity of the system, $\Delta_i=\Delta$, $G_{ij}=G$, and
$E_i=E$, the gap equation becomes
\begin{eqnarray}
|\Delta|&=& (-g) \frac{|\Delta|}{2 E} \tanh \left(\frac{\beta
E}{4}\right).
\end{eqnarray}
At zero temperature, the gap equation can be simplified as
\begin{eqnarray}
|\Delta| &=& (-g) \frac{|\Delta|}{2 E},
\end{eqnarray}
and results in a solution $|\Delta|=\sqrt{g^2-\epsilon_1^2}/2$.
The transition from a normal phase ($\Delta=0$) to a
superconducting phase ($\Delta\neq 0$) appears at a critical
coupling $g_c=-\epsilon_1$ as $g$ goes below $g_c$. These results
agree with the exact solution.

Now we turn to calculate the pair fraction. The MF ground state
can be obtained as
\begin{eqnarray}
\left| {{\psi _g}} \right\rangle  = \prod\limits_i {\left(
{{u_i}c_{i \uparrow }^\dag c_{i \downarrow }^\dag  + {v_i}}
\right)} \left| {{\rm{vac}}} \right\rangle .
\end{eqnarray}
Then the pair density matrix is of the form
\begin{eqnarray}
&&\rho _{{i_1}{\sigma _1},{i_2}{\sigma _2};{{i}_1'}{{\sigma
}_1'},{{i}_2'}{{\sigma}_2'}}^{{\rm{pair}}} = \left\langle {\hat
c_{{i_1}{\sigma _1}}^\dag \hat c_{{i_2}{\sigma _2}}^\dag {{\hat
c}_{{{i}_2'}{{\sigma}_2'}}}{{\hat c}_{{{i}_1'}{{\sigma}_1'}}}}
\right\rangle\nonumber\\
&=& {\delta _{{i_1}{i_2}}}{\delta _{{{i}_1'}{{i}_2'}}}{\delta
_{{\sigma _1}, - {\sigma _2}}}{\delta _{{{\sigma}_1'}, - {{\sigma
}_2'}}} (\delta_{\sigma_1,\uparrow}-\delta_{\sigma_1,\downarrow})
(\delta_{\sigma_1',\uparrow}-\delta_{\sigma_1',\downarrow})\nonumber\\
&\times&\left[ \delta_{i_1,i_1'} |u_{i_1}|^2+ (1-\delta_{i_1,i_1'}) u^*_{{i_1}}v^*_{{i_1'}} u_{{i_1'}}v_{{i_1}}\right]\nonumber\\
&+&
(1-\delta_{i_1,i_2})\left[{\delta _{{i_1}{i_1'}}}{\delta _{{{i}_2}{{i}_2'}}} {\delta
_{{\sigma _1},  {\sigma _1'}}}{\delta _{{{\sigma}_2},  {{\sigma
}_2'}}}\right.\nonumber\\
&-&\left. \delta_{{i_1}{i_2'}}{\delta _{{{i}_2}{{i}_1'}}}{\delta
_{{\sigma _1},  {\sigma _2'}}} {\delta _{{{\sigma}_2},
{{\sigma}_1'} }} \right].
\end{eqnarray}
In the uniform case, only the off-diagonal elements
\begin{eqnarray}
\rho _{i\sigma ,i( - \sigma );i'\sigma ,i'( - \sigma
)}^{{\rm{pair}}} =  - \rho _{i\sigma ,i( - \sigma );i'( - \sigma
),i'\sigma }^{{\rm{pair}}} = {\left| u \right|^2}{\left| v
\right|^2}\nonumber\\
\end{eqnarray}
($i\neq i'$) contribute to the macroscopic eigenvalue
$\lambda^{(0)}$ and hence the pair fraction $P$ in the large-$N$
limit\cite{Penrose56,Yang62,Leggett06,Sun09}. They are obtained as
\begin{eqnarray}
\lambda^{(0)} &\approx& 2N {\left| u \right|^2}{\left| v
\right|^2}= \frac{N}{2}\left[1-\left(\frac{\varepsilon}{E}\right)^2\right],\\
P &=& \frac{\lambda_{\rm max}-2}{N}\approx
\frac{(g^2-\epsilon_1^2)}{2g^2}.
\end{eqnarray}
The above equations work only for $g<0$ or attractive interaction.
We can see that the pair fraction $P$ rises from zero when $g<
g_c=-\epsilon_1$ (see red circles in Fig.~\ref{two level figure}),
which means that the superconducting pairing appears as the
attractive interaction becomes stronger than the critical value.
This predicted $g_c=-1$ from the pair fraction agrees with our ED
calculations. We also see a trend that the ED results approach the
MF ones as $N$ increases.

\section{$\bZ_2$ topological invariant in a class DIII chain}\label{Z2 compute}

In this Appendix, we compute $\bZ_2$ invariant for the BdG
Hamiltonian $H_p^{\rm{BdG}}$ in Eq.~(\ref{BdG H p}), which can
distinguish the topologically nontrivial and trivial phases more
rigorously. To simplify the problem, let us first perform an
$SU(2)$ transformation in spin basis
\begin{eqnarray}
\bma c^\dagger_\uparrow \\
c^\dagger_\downarrow \ema = \bma
\nu & \eta \\
-\eta^* & \nu^* \ema \bma
c'^\dagger_\uparrow \\
c'^\dagger_\downarrow \ema .
\end{eqnarray}
The unitarity of $SU(2)$ requires $|\nu|^2+|\eta|^2=1$.  After the
$SU(2)$ transformation, the pairing functions in $H^{\rm{BdG}}_p$
is given by
\begin{eqnarray}
\Delta'_\uparrow &=& \Delta_\uparrow \nu^2+\Delta_\downarrow \eta^{*2}+\sqrt{2}\Delta_{+}\nu\eta^*, \\
\Delta'_\downarrow &=& \Delta_\uparrow\eta^2+\Delta_\downarrow\nu^{*2}-\sqrt{2}\Delta_{+}\nu^*\eta, \\
\Delta'_+ &=& -\sqrt{2}\Delta_\uparrow\nu\eta+\sqrt{2}\Delta_\downarrow\nu^*\eta^*+\Delta_{+}(|\nu|^2-|\eta|^2), \\
\Delta'_- &=& \Delta_-.
\end{eqnarray}
Therefore, $\Delta_-$ is invariant under $SU(2)$ due to the
singlet pairing. Furthermore, we find
$\Delta_+^2-2\Delta_\uparrow\Delta_\downarrow$ and
$|\Delta_t|^2=|\Delta|^2+|\Delta_\uparrow|^2+|\Delta_\downarrow|^2$
also invariant under the $SU(2)$ transformation. We note that the
time-reversal constraints for the pairings in Eq.~(\ref{Trestrict
pair}) still hold under $SU(2)$ so $\Delta_-$ is real. By choosing
a proper $SU(2)$ transformation, the three triplet pairings can be
simplified as $\Delta_\uparrow=\Delta_\downarrow\equiv
\Delta_t/\sqrt{2}$ is real and $\Delta_+$ vanishes. Therefore, the
BdG Hamiltonian can be written as
\begin{eqnarray}
H^{\rm{BdG}}_p= \bma
\frac{-2t \cos p -\mu}{2} & 0 & \frac{i \sin p \Delta_t}{\sqrt{2}} & -\frac{  \cos p \Delta_-}{\sqrt{2}} \\
0 & \frac{-2t \cos p -\mu}{2} & \frac{\cos p\Delta_-}{\sqrt{2}} & \frac{i \sin p \Delta_t}{\sqrt{2}} \\
-\frac{i \sin p \Delta_t}{\sqrt{2}} & \frac{\cos p \Delta_-}{\sqrt{2}} & \frac{2t \cos p +\mu}{2} & 0 \\
-\frac{\cos p \Delta_-}{\sqrt{2}} & -\frac{i \sin p
\Delta_t}{\sqrt{2}} & 0 & \frac{2t \cos p +\mu}{2} \ema.
\nonumber\\
\end{eqnarray}
After performing a unitary transformation
\begin{eqnarray}
U=\frac{1}{2} \bma
i & -1 & -i & 1 \\
1 & -i & 1 & -i \\
-i & 1 & -i & 1 \\
-1 & i & 1 & -i \\
\ema,
\end{eqnarray}
we can simplify the BdG Hamiltonian as
\begin{eqnarray}
&& H'^{\rm{BdG}}_p = UH^{\rm{BdG}}_pU^\dagger \nonumber\\
&=& \bma
0 & 0 & A_+ e^{-i\theta_+} & 0 \\
0 & 0 & 0 & A_- e^{-i\theta_-}  \\
A_+ e^{i\theta_+} & 0 & 0 & 0 \\
0 & A_- e^{i\theta_-} & 0 & 0 \\
\ema,
\end{eqnarray}
where
\begin{eqnarray}
A_\pm(p) e^{i\theta_\pm(p)}=\frac{2t\cos p
+\mu}{2}+\frac{i}{\sqrt{2}}(\cos p \Delta_-\pm \sin p
\Delta_t).\nonumber\\
\end{eqnarray}
Similarly, the time-reversal operator under the unitary
transformation becomes
\begin{eqnarray}
\Theta'= \bma
0 & i\tau_y \\
i\tau_y & 0 \\
\ema K.
\end{eqnarray}
Solving the eigen problem in the half filling scenario, we have
two occupied eigenstates with negative energies,
\begin{eqnarray}
\ket{u^{\rm{I}}(p)} &=& \bma e^{-i \theta_+(p)/2} & 0 &
e^{i\theta_+(p)/2} & 0
\ema^T, \\
\ket{u^{\rm{II}}(p)} &=& \bma 0 & e^{-i \theta_-(p)/2} & 0 &
e^{i\theta_-(p)/2} \ema^T.
\end{eqnarray}
Furthermore, these two states are time-reversal partners
($\ket{u^{\rm{I}}(p)}=\Theta'\ket{u^{\rm{II}}(-p)},\
\ket{u^{\rm{II}}(p)}=-\Theta'\ket{u^{\rm{I}}(-p)}$).

Finally, we are able to compute the topological invariant from the
occupied states. The definition of the $Z_2$ topological invariant
in one dimension for symmetry class DIII is given
by\cite{Fu06,Budich13}
\begin{eqnarray}
P^I_o=\frac{1}{2\pi}\left[\int^\pi_0 dp \mathcal{A}_o(p)+i \ln
\left( \frac{\rm{Pf} \theta_o (\pi)}{\rm{Pf} \theta_o (0)} \right)
\right],
\end{eqnarray}
where
$\mathcal{A}_o(p)=-i(\bra{u^{\rm{I}}(p)}\partial_p\ket{u^{\rm{I}}(p)}+\bra{u^{\rm{II}}(p)}\partial_p\ket{u^{\rm{II}}(p)})$,
$\theta_o(p)$ is a matrix defined as
$\theta_o^{\alpha\beta}(p)=\bra{u^{\alpha}(p)}\Theta'\ket{u^\beta
(-p)}$ and Pf denotes the Pfaffian. When particle-hole symmetry is
present, $P^I_o$ is quantized and its value (mod $1$) describes
topology in 1D time-reversal superconductors (0 trivial and 1/2
nontrivial). In our case,
\begin{eqnarray} P^I_o=\frac{i}{2\pi}
\ln \left(
\frac{\cos(\frac{\theta_-(0)-\theta_+(0)}{2})}{\cos(\frac{\theta_-(\pi)-\theta_+(\pi)}{2})}
\right),
\end{eqnarray}
where
\begin{eqnarray}
\theta_-(0)-\theta_+(0) &=& 0,\\
\theta_-(\pi)-\theta_+(\pi) &=&
\begin{cases}
0, \quad\quad \rm{as}\ \frac{4\Delta_t^2 t^2}{\Delta_t^2+\Delta_-^2}<\mu^2, \\
-2\pi, \ \ \rm{as}\  \frac{4\Delta_t^2
t^2}{\Delta_t^2+\Delta_-^2}>\mu^2,
\end{cases}
\end{eqnarray}
Therefore, when $\frac{4\Delta_t^2
t^2}{\Delta_t^2+\Delta_-^2}>\mu^2$, $P^I_o=1/2$ corresponds to a
topologically nontrivial phase, which is consistent with the
topological region in Eq.~(\ref{eqn:topo region}) with $SU(2)$
invariant $\Delta_-$ and  $\Delta_t^2$.

\section{Trends of the pair fraction in a Hubbard chain with onsite interaction}\label{sec:HubbardU}

\begin{figure}
\includegraphics[width=7cm]{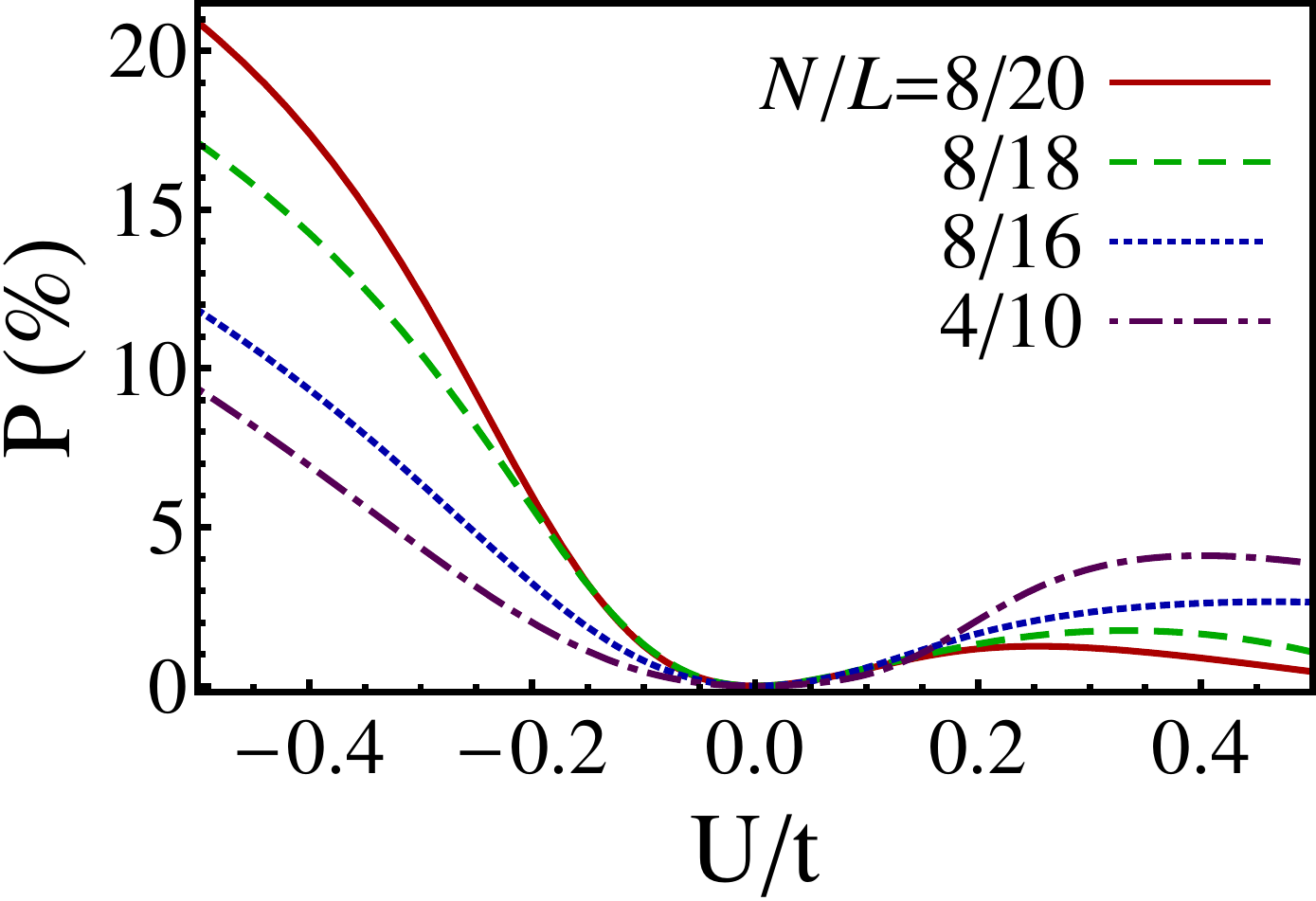}
\caption{(Color online) The relative pair fraction $P$ vs on-site
interaction $U$ in a Hubbard chain with various numbers of
particles $N$ and sites $L$. The red solid, green dashed, blue
dotted, and purple dot-dashed curves represent the cases of
$(N,L)=(8,20)$, $(8,18)$, $(8,16)$, and $(4,10)$, respectively. In
the attractive interaction region ($U<0$), increasing size at
fixed $N$ (as the system dilutes) and at fixed $N/L$ (extends)
coincides with positive and increasing $P$, so the system is
regarded as a stable pairing state or a superconducting state in
the thermodynamic limit. Such trends do not hold in the repulsive
interaction region ($U>0$). Therefore, we obtain the transition
point at $U=0$, consistent with the solution from the BCS gap
equation (see text).} \label{fig:trend in HubbardU}
\vspace{-0.5cm}
\end{figure}

In this Appendix, we show that the three signatures of the
relative pair fraction $P$ [defined in Eq.~(\ref{eqn:P})], (i)
being positive, (ii) increasing as the system extends, and (iii)
increasing as the system dilutes, which are used in
Sec.~\ref{sec:ED} to identify a stable pairing state in our
extended Hubbard chain, also apply to the original Hubbard chain
with only on-site interaction. The Hamiltonian of the original
Hubbard model has the same form as Eq.~(\ref{eqn:Ham1}) with the
nearest-neighbor couplings $V$ and $J$ vanishing. In this case,
the on-site interaction $U$ can induce only the singlet pairing in
the system.

First, we perform exact diagonalization on a finite-size setup
similar to that in Sec.~\ref{sec:ED}, with the same noninteracting
terms and the interacting terms replaced by the on-site
interaction. Figure \ref{fig:trend in HubbardU} shows $P$ as a
function of $U$ at various particle numbers
$N_\uparrow=N_\downarrow=N/2$ and sizes $L$ of the system. In the
attractive region ($U<0$), comparing the cases of $(N,L)=(8,16)$,
$(8,18)$, and $(8,20)$ (blue dotted, green dashed, and red solid
curves, respectively), we see positive and increasing $P$ as the
system dilutes. Comparing the cases of $(N,L)=(4,10)$ and $(8,20)$
(purple dot-dashed and red solid curves, respectively), we see
positive and increasing $P$ as the system extends at a fixed
density. In the repulsive region ($U>0$), although $P$ can be
positive, the other signatures disappear. As the trends persist
toward the thermodynamic limit, we expect that $P$ approaches a
finite value, indicating a stable pairing or superconducting
state, at $U<0$ and $0$, indicating a normal state, at $U>0$. The
transition point is thus $U=0$.

Second, we apply the same mean-field treatment as in
Sec.~\ref{sec:MF} and obtain the BCS gap equation,
\begin{eqnarray}
\frac{{\Delta _ - ^*}}{U} =  - \frac{{\Delta _ - ^*}}{{\beta
L}}\sum\limits_{p,n} {\frac{1}{{\omega _n^2 + {{\left( {E_p^0 -
\mu } \right)}^2} + {{\left| {{\Delta _ - }} \right|}^2}}}},
\end{eqnarray}
where $E^0_p$ is the single-particle energy spectrum. The gap
equation has nonzero solutions if $U<0$ and the only solution of
$\Delta_-=0$ if $U>0$. These also indicate a transition point at
$U=0$. Therefore, with the use of the three signatures, the
exact-diagonalization results agree with those from the mean-field
treatment.

\section{SU(2) symmetry and eigenvalues of pair density matrix}\label{SU2}
In this Appendix, we show that the three triplet blocks of the
pair density matrix in Eq.~(\ref{eqn:PDM}) are identical under
$SU(2)$ symmetry and hence have the same set of eigenvalues.
Provided that there is a unique ground state subject to our
Hamiltonian under $SU(2)$ symmetry, it should also be invariant
under the $SU(2)$ transformation. In addition, our Hamiltonian
commutes with the total spin $\hat S_z$ ($=\hat N_\uparrow - \hat
N_\downarrow$) of the system, so $S_z$ is a good quantum number
for the unique ground state. In other words, any spin-flip
operator that changes $S_z$ should vanish when sandwiched by the
ground state.

A general form of the pair density matrix is block-diagonalized
with two intraspin blocks and one interspin block, due to the
$S_z$ conservation. The interspin trplet block can be further
separated from the singlet one after a proper transformation. As a
result, the matrix elements of the three triplet blocks that
correspond to the same spatial coordinate $\{i,j\}$ can be written
respectively as
\begin{eqnarray}
m_\uparrow &=& \langle \hat c^\dagger_{j \uparrow} \hat c^\dagger_{i \uparrow}\hat c_{i \uparrow}\hat c_{j \uparrow}\rangle,        \\
m_\downarrow &=&\langle \hat c^\dagger_{j \downarrow}\hat c^\dagger_{i \downarrow}\hat c_{i \downarrow}\hat c_{j \downarrow} \rangle,  \\
m_+ &=&\frac{1}{2}\langle(\hat c^\dagger_{j \downarrow }\hat
c^\dagger_{i \uparrow}+\hat c^\dagger_{j \uparrow }\hat
c^\dagger_{i \downarrow})(\hat c_{i \uparrow}\hat c_{j \downarrow
}+\hat c_{i \downarrow}\hat c_{j \uparrow }) \rangle.
\end{eqnarray}
Performing an $SU(2)$ transformation,
\begin{eqnarray} \hat
c_\uparrow=\frac{1}{\sqrt{2}}(\hat c'_\uparrow+c'_\downarrow),
\quad \hat c_\downarrow=\frac{1}{\sqrt{2}}(-\hat c'_\uparrow+\hat
c'_\downarrow),
\end{eqnarray}
we obtain a relation between the
matrix elements in the original and the new spin basis as
\begin{eqnarray}
m_\uparrow &=& \frac{1}{4}(m'_\uparrow+m'_\downarrow+2m'_+), \\
m_\downarrow &=& \frac{1}{4}( m'_\uparrow + m'_\downarrow + 2m'_+ ), \\
m_+ &=& \frac{1}{2}(m'_\uparrow+m'_\downarrow),
\end{eqnarray}
which immediately shows $m_\uparrow=m_\downarrow$. Since each
matrix element is a physical observable (two-body correlation),
which should be the same $SU(2)$ invariant as the Hamiltonian, we
have
\begin{eqnarray}
m'_\uparrow=m_\uparrow , \ m'_\downarrow=m_\downarrow,\ m'_+=m_+.
\end{eqnarray}
Combining these relations, we obtain
\begin{eqnarray}
m_\uparrow=m_\downarrow=m_+.
\end{eqnarray}
The result is valid for every spatial coordinate $\{i,j\}$, so the
three triplet blocks are identical.


\begin{thebibliography}{99}
\bibitem{tuningV}
An atomic Fermi gas with two hyperfine states can be treated as an
effective spin-half system. In such system, it is possible to
differently tune the interactions $V_{\uparrow \uparrow}$,
$V_{\downarrow \downarrow}$ and $V_{\uparrow \downarrow}$ via the
tuning of the Feshbach resonance (see Ref.~\onlinecite{Chin10} for
details).
\bibitem{numberconserving}
This ansatz is difference from but as physically effective as the
well-known BCS form that breaks $U(1)$ symmetry. See detailed
discussions in Ref.~\onlinecite{Leggett06}.
\end{thebibliography}
\end{document}